%% file: main.tex
\documentclass[sigconf]{acmart}
\AtBeginDocument{%
  }

\copyrightyear{2026}
\acmYear{2026}
\setcopyright{cc}
\setcctype{by}
\acmConference[ASSETS '26]{The 28th International ACM SIGACCESS Conference on Computers and Accessibility}{October 25--28, 2026}{Vila Nova de Gaia, Portugal}
\acmBooktitle{The 28th International ACM SIGACCESS Conference on Computers and Accessibility (ASSETS '26), October 25--28, 2026, Vila Nova de Gaia, Portugal}
\acmDOI{10.1145/3797867.3829041}
\acmISBN{979-8-4007-2521-0/2026/10}

\newcommand{\change}[1]{{\textcolor{black}{#1}}}
\newcommand{\revise}[1]{{\textcolor{black}{#1}}}

\usepackage{booktabs,tabularx}
\usepackage{makecell}

\begin{document}

\title[Co-designing Protection Mechanisms with PWD in Social VR]{Safety vs. Social Image: Co-Designing Protection Mechanisms Against Ableist Harassment with People with Disabilities in Social Virtual Reality}


\author{Kexin Zhang}
\orcid{0009-0009-4078-8780}
\affiliation{%
   \department{Department of Computer Sciences}
   \institution{University of Wisconsin-Madison}
   \city{Madison}
   \state{Wisconsin}
   \country{USA}
}
\email{kzhang284@wisc.edu}

\author{Daniel Killough}
\orcid{0009-0002-2623-0528}
\affiliation{%
   \department{Department of Computer Sciences}
   \institution{University of Wisconsin-Madison}
   \city{Madison}
   \state{Wisconsin}
   \country{USA}
}
\email{dkillough@wisc.edu}

\author{Xinran Adeline Li}
\orcid{0009-0000-4822-786X}
\affiliation{%
   \department{Department of Computer Sciences}
   \institution{Johns Hopkins University}
   \city{Baltimore}
   \state{Maryland}
   \country{USA}
}
\email{xli436@jh.edu}

\author{Yaxing Yao}
\orcid{0009-0008-7900-9265}
\affiliation{%
   \department{Department of Computer Sciences}
   \institution{Johns Hopkins University}
   \city{Baltimore}
   \state{Maryland}
   \country{USA}
}
\email{yaxing@jhu.edu}

\author{Yuhang Zhao}
\orcid{0000-0003-3686-695X}
\affiliation{%
   \department{Department of Computer Sciences}
   \institution{University of Wisconsin-Madison}
   \city{Madison}
   \state{Wisconsin}
   \country{USA}
}
\email{yuhang.zhao@cs.wisc.edu}

\renewcommand{\shortauthors}{Zhang et al.}
\newcommand{\kexin}[1]{{\color{magenta} (Kexin: #1)}}
\newcommand{\yaxing}[1]{{\color{orange} (YY: #1)}}
\newcommand{\daniel}[1]{{\color{yellow} (DK: #1)}}
\begin{abstract}
People with disabilities (PWD) increasingly use avatars to express disability identities in social virtual reality (VR), but greater visibility also invites targeted harassment. Existing safety features are often insufficient, overlooking PWD's experiences and needs. To address this gap, we co-designed protection mechanisms with 11 PWD to reveal their values and needs. Our research employed a social lens to interpret harassment behaviors and protection mechanisms. Inspired by Hall’s Proxemics Theory that interpersonal distances indicate social intent and boundaries, we divided social VR spaces into four proxemic zones (\textit{Intimate}, \textit{Personal}, \textit{Social}, and \textit{Public}) and used them to structure our protection mechanism co-design. We also provided different protection mechanism probes (\textit{Inform}, \textit{Educate}, \textit{Consent}, and \textit{Defense}) to elicit participant preferences. Our study highlighted the role of social proximity in shaping PWD’s harassment perception and protection preferences and revealing PWD's unique social values and needs (e.g., managing harassment with optimism and resilience, prioritizing social image over safety). We proposed design recommendations for protection mechanisms that protect PWD while maintaining their desired social images.

\end{abstract}






\maketitle

\section{Introduction}
\input{sections/1-introduction}

\section{Related Work}
\input{sections/2-related_work}

\input{sections/3-method}

\section{Findings}
\input{sections/4-findings}
\section{Discussion}
\input{sections/5-discussion}

\section{Conclusion}
\input{sections/6-conclusion}

\begin{acks}
\revise{We thank all anonymous participants for their efforts and valuable feedback. This work was supported in part by the National Science Foundation under Grant No. IIS-2328182, No. IIS-2608524, and No. CNS-2442221.}
\end{acks}

\bibliographystyle{ACM-Reference-Format}
\bibliography{references}

\appendix
\newpage
\section{Appendix}
\input{sections/7-appendix}


\end{document}

%% file: sections/1-introduction.tex
People with disabilities (PWD) have shown increasing presence in social virtual reality (VR), which enables multiple users to meet and interact as embodied avatars \cite{zhang_2022}. 
Prior work has found that PWD use their embodied avatars to disclose disability identities and spark meaningful social interactions, such as conversations about identity and disability awareness \cite{zhang_2023_diary, angerbauer2024_isit?, zhang_2022, Gualano_invisible_full2024, mack2023towards}. However, expressing disability identities in social VR also makes PWD more vulnerable to targeted harassment, such as repeated use of ableist terms, social exclusion in multiplayer games, and unwanted interactions with PWD's avatars (e.g., grasping their virtual cane) \cite{zhang_2023_diary, angerbauer2024_isit?}.

Despite the growing awareness of disability-targeted harassment in social VR \cite{zhang_2023_diary, angerbauer2024_isit?}, it remains unclear how to effectively protect PWD while enabling them to freely express their disabilities. Mainstream social VR platforms have implemented various safety features, such as muting, blocking, reporting, and establishing personal boundaries \cite{abhinaya2024enabling, sharma_2022_personalboundary}, but these features often employ a reactive approach that only responds to harassment after it occurs \cite{liao2025_proactive_safety}. In 2022 Meta released the ``Personal Boundary'', a more preventive feature to keep a minimum distance between avatars and limit unwanted interactions \cite{Meta2022PersonalBoundary}; however, it blocks all forms of social interactions, even though some PWD welcome engagement with their avatars with disabilities upon consent 
\cite{zhang_2023_diary}. 
No research has investigated PWD's needs and preferences for protection mechanisms, leaving important questions unaddressed: \textbf{How do PWD's harassment perceptions shape their needs and preferences for protection mechanisms? How can we design effective protection mechanisms that center on PWD's experiences?}


To answer these questions, \change{we conducted a co-design study} 
with 11 PWD, \revise{involving them as co-creators and experts} to reveal their underlying values and needs for protection mechanisms. 
\revise{We used Hall’s Proxemic Theory to structure our co-design within the broad design space of social VR protection mechanisms, as proximity is key to establishing social boundaries and behaviors in both physical and virtual worlds \cite{hall_1966_hidden, restrainingorder_1985, Low2019Spikes, ji2022vrbubble}, thus impacting the harassment perception, perceived safety, and protection choices.}
\revise{Following Hall's Proxemic Theory \cite{hall_1966_hidden}, we} divided the space surrounding an avatar into four distance layers (\textit{Intimate}, \textit{Personal}, \textit{Social}, and \textit{Public}) to structure the co-design. For each social distance, we illustrated different harassment scenarios to understand participants' harassment perceptions. We presented four protection mechanisms---\textit{Inform}, \textit{Educate}, \textit{Consent}, and \textit{\revise{Defense} Mechanism}---as design probes to elicit their design preferences and needs.

Our study \change{revealed that proximity operated as a rich socio-spatial signal through which PWD read the target, intent, and severity of disability-targeted harassment.} 
Beyond proximity, PWD used behavioral cues such as the perceived effort invested in harmful behaviors and bystanders' reactions to evaluate harm severity. Moreover, our findings captured PWD's unique values that shaped their needs and preferences for protection mechanisms. Participants with optimistic, resilient mindsets viewed harassment as opportunities for disability education and advocacy, while those with risk-averse mindsets perceived harassment as threats that required systematic intervention and norm establishment. Interestingly, we noted that some PWD prioritized maintaining a desired social image over personal safety, which further shaped their needs and preferences for protection mechanisms (e.g., preferred a less confrontational approach to maintain a sociable self-image). 
Based on our findings, we proposed design implications for protection mechanisms that maintain both PWD's safety and desired social images in social VR. 

Our research makes two main contributions to the \change{ASSETS} community. First, we contribute the first \change{co-design study that} 
uncovered PWD's harassment perception, values, and needs that shape their protection mechanism choices. Second, we proposed rich design recommendations for protection mechanisms that can effectively protect PWD while maintaining their desired social images.

%% file: sections/2-related_work.tex
\change{Our work investigates PWD's unique needs and preferences for protection mechanisms in social VR. We situate this work within existing literature on \revise{Hall's Proxemics Theory,} social stigma and the protection strategies PWD develop in everyday life, as well as disability harassment and the current landscape of mitigation strategies in social VR.} 

\revise{\subsection{Proximity-based Interactions in Physical and Virtual Worlds}}
\revise{In a shared space, interpersonal distance plays a critical role in shaping social interactions. Edward Hall coined the term \textit{Proxemics} to describe how people use interpersonal distance to interpret and mediate their interactions with others \cite{hall_1966_hidden}. Hall divided the space surrounding a person into four proxemic zones, each governed by distinct social norms: intimate (less than 1.5 feet) for touch and close contact, personal (1.5 to 4 feet) for conversations among friends, social (4 to 12 feet) for interactions with acquaintances, and public (12 to 25 feet) for formal occasions such as public speaking. Because each zone specifies who may enter and under what conditions, proximity naturally functions as a boundary-setting and self-protection mechanism in the physical world. People reposition or physically distance themselves when others intrude into their proxemic zones \cite{Greenberg2011ProxemicInteractions}, and society also formalizes this protective logic. For example, courts may issue restraining orders that require abusers to keep a fixed distance from protected individuals, turning physical separation into a legal safety mechanism \cite{restrainingorder_1985}.} 

\revise{Notably, proximity-based norms have carried over into virtual worlds \cite{ji2022vrbubble, bailenson2003interpersonal}. Bailenson et al. found that users maintained greater distances from virtual humans who faced or gazed at them and moved farther away when virtual humans entered their personal space \cite{bailenson2003interpersonal}. Ji et al. developed \textit{VRBubble}, which divided surrounding avatars into distance-based zones and provided corresponding audio feedback to support avatar awareness for people with visual impairments \cite{ji2022vrbubble}. These studies show the potentials of proximity in reflecting changing social boundaries and supporting context-aware system responses in virtual worlds. However, how proximity shapes perceptions of harassment and preferences for protection mechanisms remains underexplored, especially for PWD in social VR. This gap motivates our investigation of how protection needs may vary across proxemic zones. Building on this theoretical lens, our work uses Hall's four proxemic zones to structure the design of protection mechanisms, examining how protection needs may vary across interpersonal distance in social VR.}

\subsection{\change{Combating Ableist Harassment in Everyday Social Life}}
PWD encounter multiple barriers that largely restrict their involvement in everyday social life, leading to isolation and loneliness \cite{emerson2021loneliness, hall2009social, tough2017social}. These barriers are systematically rooted in inaccessible environments and various forms of social stereotypes, such as negative perceptions associated with assistive technology use \cite{howard2022ATuse, eisenberg2015vulnerable} and childhood segregation from peers \cite{daley2018stigma}. 
Media representations further stigmatize PWD by misrepresenting them as objects of pity, inspirational superheroes, or charity recipients \cite{quayson2007aesthetic, young2014m, mack2024t2i}. 
Collectively, PWD have limited access to social activities while being victims of ableism 
\cite{heung2022nothing, heung2025ignorance}.

In response, PWD have developed diverse strategies to protect their well-being and social image across physical and digital social settings \cite{stetten2019SupportGroups, Low2019Spikes, lingsom2008invisibleImpairments, miron2023onlinedating, heung2025ignorance}. For example, in physical public spaces, wheelchair users attach spikes to their handles or remove push bars entirely to deter strangers from pushing them without consent \cite{Low2019Spikes}. On 2D social media, PWD carefully manage visibility by curating when and how disability appears in posts, bios, and photos \cite{sannon2023disability}. They also leverage platform moderation tools like blocking, muting, and keyword filtering to gate ableist content \cite{heung2025ignorance}. Disability advocates further respond to hate by creating educational posts that explain why a comment or incident is ableist, attempting to turn the harassment into disability education for their followers~\cite{heung2024vulnerable, sannon2023disability}.

While PWD's protection strategies have been well documented in real life and 2D social media, how PWD protect themselves in social VR remains underexplored. Our work aims to address this gap by investigating how PWD's values may carry into social VR and what new protection mechanisms they need in this emerging social space.



\subsection{Ableist Harassment in Social VR}
Social VR has grown in popularity, providing an immersive virtual space where users can meet and interact through embodied avatars \cite{freeman2020my, zhang_2022}. 
By rendering full-body movements and mediating both verbal and non-verbal cues \cite{maloney2020talking, moustafa2018longitudinal, zamanifard2019togetherness}, social VR 
foster a strong sense of embodiment and co-presence, making many users view avatars as proxies and curating them to meet their social goals \cite{freeman2020my, morris2023don, zhang_2025_guidelines}. \change{For PWD, these social VR affordances could be particularly valuable, which bypass physical and social barriers that PWD often encounter in real life \cite{howard2022ATuse, emerson2021loneliness} by enabling participation in shared communities \cite{tough2017social} and supporting identity expression 
through avatar customization \cite{zhang_2022, zhang_2023_diary, 
Gualano_invisible_full2024}.}


Recent research attention has examined PWD’s experiences of using avatars with disability signifiers in social VR \cite{zhang_2022, Gualano_invisible_full2024, zhang_2023_diary}. While many PWD preferred to represent their disabilities through avatars, doing so can expose them to ableist harassment in social VR \cite{zhang_2023_diary, angerbauer2024_isit?}. Zhang et al. \cite{zhang_2023_diary} conducted a two-week diary study with 10 PWD to investigate the impacts of using avatars with disability signifiers (e.g., a virtual wheelchair, a virtual cane). They found that using avatars with disability signifiers triggered multiple types of disability-targeted harassment, including verbal harassment (e.g., repetitive use of ableist language, being described as inferior or incapable), physical harassment (e.g., non-consensual actions like pushing a virtual wheelchair), and environmental harassment (e.g., mimicking one's disability through avatar with stereotypical portraits of disabilities). Angerbauer et al. \cite{angerbauer2024_isit?} expanded prior work by conducting a one-week diary study to understand the experiences of 26 PWD when using avatars with both visible and invisible disability signifiers (e.g., avatar with a sunflower patch representing chronic health conditions). They revealed that, even with the deliberate effort of designing signifiers for positive disability representation, many design choices still result in negative user experiences such as ableist hate. These works highlight the urgent need for effective protection mechanisms that empower PWD to freely express themselves while safely using social VR.

\subsection{Harassment Mitigation in Social VR}
Despite the growing awareness of embodied harassment, how to effectively mitigate its harm is a challenging task due to unestablished social norms in social VR \cite{blackwell2019harassment, Schulenberg_AImod_2023}. 
Multiple efforts from academia and the social VR industry have explored solutions to mitigate harmful behaviors in social VR \cite{weerasinghe2025beyond, fiani2024pikachu, fiani_bigbuddy_2023, Schulenberg_AImod_2023, vrchat_safety_trust_2025}. In this section, we review mainstream harassment mitigation strategies and discuss their benefits and limitations. 

\textbf{Community Guidelines} are the baseline for governing social VR \cite{gray2024_book, lazerson_secure_2022}. They set clear expectations of proper behaviors and serve as the foundation for proactively prohibiting harmful content \cite{lazerson_secure_2022}. Multiple mainstream social VR platforms have published community guidelines, such as \textit{Rec Room}'s Code of Conduct \cite{RecRoom_CodeOfConduct} and \textit{VRChat}'s Community Guidelines \cite{VRChat_Community_Guidelines_2024}. However, enforcement of community guidelines is critical to ensure its effectiveness. If no enforcement exists, a lack of clear outcomes can create the perception that harmful behavior has no consequences, which may encourage further misconduct \cite{gray2024_book, fiani_bigbuddy_2023}.

\textbf{Platform-provided Safety Features} are commonly used to mitigate harassment, typically allowing users to mute, block, and report harassers \cite{Chen2025DemocraticModeration, zheng2023understanding, freeman2022disturbing}. 
Although widely used, prior work has identified several limitations of these approaches. First, they place the burden on users who may be new to social VR or unfamiliar with effective protection strategies  \cite{freeman2022disturbing, Schulenberg_AImod_2023}. In rapidly escalating situations, users may not have time to activate these features for effective self-protection. Second, some safety features can be misused to create harmful behaviors, such as vote-kick systems being weaponized against marginalized users \cite{zheng2023understanding, Chen2025DemocraticModeration}. Lastly, most safety features are reactive (e.g., mute and block), failing to prevent harm from occurring in the first place \cite{liao2025_proactive_safety}. 

Some social VR platforms (e.g., \textit{Meta Horizon}, \textit{VRChat}) have started implementing more proactive methods, such as the ``Personal Boundary''\cite{Meta2022PersonalBoundary}, which makes nearby avatars invisible if they get too close to a user. However, such features may disrupt immersive experiences by blocking all social interactions, including positive ones. This trade-off can be especially costly for PWD, as prior work found that their avatars' disability signifiers may invite meaningful social interactions \cite{zhang_2023_diary}. 


\textbf{AI-based Moderation} has recently emerged as a promising solution to automatically detect, classify, and respond to harassment behaviors in social VR \change{\cite{Schulenberg_AImod_2023, freeman_comforting_2025, xu2024safe, lee2025harassguard, fiani2024pikachu}. A growing body of HCI literature has empirically investigated people's attitudes and needs towards AI-based moderation \cite{schulenberg_2023_birdcage, fiani2024pikachu, freeman_comforting_2025, fiani_bigbuddy_2023}. For example, Schulenberg et al. \cite{Schulenberg_AImod_2023} interviewed 39 social VR users and found that people perceived AI moderation as offering large-scale and consistent judgment, but it often misreads context and amplifies bias. 
Freeman et al. \cite{freeman_comforting_2025} focused on gender-based harassment and interviewed 20 female users to envision AI support for women in social VR. Participants envisioned four AI companion roles: accessible, informational, emotional supportive, and protective, while cautioning that such companions need to preserve user agency and avoid becoming new tools for surveillance or harassment.}

\change{On the technical side, researchers have leveraged multi-modal AI to detect harassment in social VR \cite{wang2024hardenvr, lee2025harassguard, xu2024safe}. For example, Xu et al. \cite{xu2024safe} implemented \textit{Safe Guard}, an LLM agent that detects hate speech in real-time voice-based interactions in VRChat; Lee et al. \cite{lee2025harassguard} deployed \textit{HarassGuard}, which used a vision-language model to capture physical harassment; Wang et al. \cite{wang2024hardenvr} developed a technical framework named \textit{HardenVR} to distinguish physical harassment behaviors (e.g., slapping, grabbing, pushing) from avatar motion and controller data by jointly modeling users' actions and their spatial-temporal relationships.}

\change{While current AI-powered systems can detect when harassment happens, they do not effectively translate that detection into well-designed protection mechanisms (i.e., what the system should do next to protect users). They also focused on generic harassment without considering PWD's unique experiences in social VR. We aim to fill this gap by drawing design insights on AI-powered protection mechanisms that center on PWD's needs and preferences.}

%% file: sections/3-method.tex
\section{Method} 
The goal of this study is to design protection mechanisms that empower PWD to freely express themselves while safely using social VR. With this goal in mind, we conducted a co-design study with 11 participants to thoroughly explore the factors that shaped PWD's harassment perceptions, as well as their preferences and needs for effective protection mechanisms. This study was approved by the Institutional Review Board (IRB) at our university. 


\subsection{Participants} 
We leveraged multiple channels for participant recruitment, including the mailing lists of non-profit disability organizations (e.g., Association for Autism and Neurodiversity, the United Spinal Association), referrals from recruited participants, and our university’s research forum. Interested participants completed a screening survey to report their age, disability conditions, social VR experiences, and experiences with disability-targeted harassment in social VR. Eligible participants must (1) be over 18 years old, (2) identify as having disabilities, and (3) have experience using social VR. 
\revise{We prioritized but did not require prior harassment experiences as we demonstrated simulated examples in the co-design activity.}
We limited our recruitment to individuals who spoke English. 

As a result, we recruited 11 participants (5 female, 4 non-binary, and 2 male), aged from 18 to 64 ($mean$ = 29, $SD$ = 13). Our participants covered a wide range of disabilities, ranging from mobility disabilities (e.g., genetic brittle bone condition), chronic health issues (e.g., chronic kidney disease, chronic pain), neurodiversity (e.g., autism, ADHD), to mental health conditions (e.g., depression, anxiety). All participants had prior social VR experience: seven had used social VR for 2–4 years, and four for a few hours to 1 year. Their varied experiences enabled us to compare needs and preferences across novice and experienced social VR users. In addition, five participants (P1, P2, P6, P8, P11) reported experiencing disability-targeted harassment in social VR (e.g., ableist slurs, mockery of disability, stereotypical ``disabled'' avatars), two reported witnessing such incidents when using social VR (P5, P9), and three reported experiencing disability-targeted harassment in real life (P3, P4, P10). We enumerate participants' demographic information and experiences in Table \ref{tab: demographics}.


\input{sections/tables/demographics}

\subsection{Apparatus}
To better understand the impact of proximity on PWD's harassment perception and preferences of protection mechanisms, we contextualize participants in embodied VR space by building a virtual environment with four proxemic zones visualized, following Hall's proxemics theory \cite{hall_1966_hidden}. Within the environment, we displayed three \change{examples} of disability-targeted harassment at each zone to systematically assess perception changes and demonstrated four design probes to guide the co-design of protection mechanisms. We implemented the virtual space using A-Frame\footnote{A-Frame. https://aframe.io/}, a framework that renders 3D spaces online for ease of deployment across various XR platforms. 
We describe the details below.

\subsubsection{Virtual Environment with Proximity Visualization} \label{sec: proximity}
We created a virtual environment to \change{represent a} 
social VR scene---a cafe lounge with an open field so that participants can freely explore both indoor and outdoor virtual spaces. We also included open seating areas and some non-player avatars in the background to cultivate a more lively social environment \change{and to better contextualize the proposed scenario to participants} (Figure \ref{fig: study apparatus}A).

To \change{illustrate} avatar-mediated interactions in social VR, we rendered two key, drivable avatars in this virtual environment: one avatar represented the participant's avatar with one or more disability signifiers (e.g., virtual wheelchair, cane, invisible disabilities symbols), and we personalized the avatar design based on each participant's disability; another avatar \change{served as an example harasser for demonstrating different types of harassment.} 
Both avatars are designed to be genderless and \change{ethnically} ambiguous to eliminate potential stereotypes. 

We followed Hall's proxemics theory \cite{hall_1966_hidden} to divide the virtual space into four interpersonal layers and visualize the proximity between the two avatars: \textbf{\textit{Intimate Distance}} (within 1.5 feet) for close interactions where avatars may touch or collide; \textbf{\textit{Personal Distance}} (1.5 to 4 feet) for casual conversations between friends; \textbf{\textit{Social Distance}} (4 to 12 feet) for interactions with less familiar acquaintances or small groups; and \textbf{\textit{Public Distance}} (12 to 25 feet \change{and beyond}) for formal settings like public speaking, where avatars can be seen or heard but do not typically engage in direct interaction. 
We visualized these distances via four concentric circles on the floor, centering on the participant's avatar, with radii of 1.5 feet, 4 feet, 12 feet, and 25 feet respectively. The distance visualization was attached to the participant's avatar to give them a consistent sense of proximity when exploring and interacting with others in the environment. \revise{Following a common visualization of Hall's Proxemics Theory \cite{hall_1966_hidden, deRosa2021Covid}}, we color-coded the four distances for easier recognition: \textit{Intimate} as a red ring, \textit{Personal} as an orange ring, \textit{Social} as a yellow ring, and \textit{Public} as a green ring (Figure \ref{fig: study apparatus}A).

\subsubsection{Three Examples of Disability-targeted Harassment} \label{sec: harassment scenarios}

\begin{figure*}
     \centering
     \includegraphics[width=\linewidth]{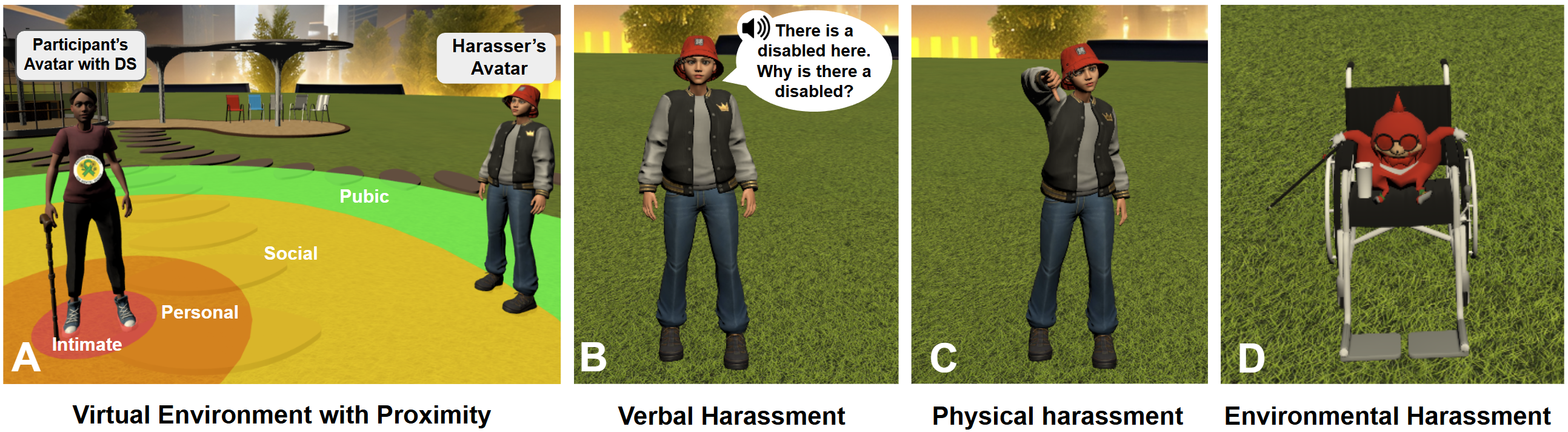}
     \caption{Study Apparatus. We created a virtual environment to \change{represent a social VR scenario with embodied proximity and}
     rendered two avatars for avatar-mediated interactions: one represented the participant's avatar with customized disability signifier(s) (P4's as shown); the other represented the harasser's avatar (A). We also 
     \change{presented three examples of disability-targeted harassment defined by prior works, including}
     verbal harassment (B), physical harassment (C), and environmental harassment (D).} 
     \Description{Figure 1. Four-panel figure illustrating the study’s VR harassment scenarios and proxemic zones. (A) A virtual outdoor scene with color-coded proximity rings around a participant avatar with a visible disability signifier (DS), labeled Intimate, Personal, Social, and Public; a second avatar labeled “Harasser’s Avatar” stands farther away. (B) Example verbal harassment, showing the harasser avatar speaking a derogatory, disability-targeted comment in a speech bubble. (C) Example physical harassment, showing the harasser making a thumb-down gesture toward the target (e.g., pointing/approaching as if to invade space or touch). (D) Example environmental harassment, showing a Ugandan knuckles avatar with overly exaggerated assistive technology use, representing a meme avatar with mocking intention.
}
     \label{fig: study apparatus}
\end{figure*}

\revise{To elicit how participants' harassment perception shifts across proximity in a 3D space and the downstream protection preferences, we presented three common examples of disability-targeted harassment in a proximity-encoded virtual environment \cite{zhang_2023_diary, blackwell2019harassment}:} 
(1) \textit{\textbf{Verbal harassment}}, where the harasser avatar said \textit{``There is a disabled here, why there is a disabled?''} 
 (Figure \ref{fig: study apparatus}B); (2) \textit{\textbf{Physical harassment}}, where the harasser avatar pointed at and conducted thumbs-down gestures 
 toward the avatar with disability (Figure \ref{fig: study apparatus}C); when a participant’s avatar had signifiers of assistive devices (i.e., virtual wheelchair, cane), we added pushing and snatching gestures to illustrate non-consensual physical interaction with PWD's disability signifiers \cite{zhang_2023_diary}; 
(3) \textit{\textbf{Environmental harassment}}, where the harasser avatar switched to a meme avatar\footnote{Meme avatar in VRChat: \url{https://vrcmods.com/item?id=6451}} seated in a wheelchair and holding a cane to mimic disability in a dehumanized manner (Figure \ref{fig: study apparatus}D). 
We recognized the concerns that demonstrating these examples may lead to mental stress of participants. We \revise{reflected on ethical considerations and} risk handling strategies in Section \ref{Ethics}. 

\subsubsection{Four Design Probes of Protection Mechanisms} \label{sec: design probes}
We also designed and prototyped four protection mechanisms as probes to explore PWD's preferences of safety features at different proximity layers. These design probes served as a lightweight demonstration of potential protection mechanisms we envisioned \cite{hutchinson2003probes}, allowing us to obtain early user feedback and derive design recommendations. 

To ensure a comprehensive coverage of protection mechanisms as design probes, we conducted a systematic literature review of harassment and mitigation efforts in social VR to synthesize existing protection mechanisms as well as identifying key factors that cause harassment to inspire novel designs.
Our literature search included both general query on Google Scholar and more targeted search on representative HCI and AR/VR venues (e.g., ACM CHI, ASSETS, CSCW, IEEE VR, ISMAR). We used the following keywords: ``online harassment,'' ``safety risks,'' ``safety designs,'' ``mitigation strategies,'' ``protection mechanisms,'' combined with ``people with disabilities'' and ``social virtual realities (or VR).'' We filtered the searched results by manually examining each paper and removing irrelevant and \revise{duplicate papers}. 
In the filtering process, we also identified papers discussing harassment and mitigation of other minoritized identities, such as women \cite{creepy_Schulenberg_2023, freeman_comforting_2025} and children \cite{fiani_bigbuddy_2023}. We reviewed these paper to comprehensively understand the broader landscape of protection strategies to inspire our design. 

As a result, we collected a list of 11 papers focused on social VR harassment and safety designs. We identified four key design strategies for protection mechanisms, including: (1) signal one’s own identities and needs to communicate interaction preferences to others \cite{zhang_2023_diary, liao2025_proactive_safety}; (2) set clear expectations and rule enforcement in a virtual world \cite{zheng2023understanding, fiani2024pikachu, sabri2023challenges}; (3) provide warnings before conflict escalations and use consent to reclaim personal boundaries \cite{schulenberg_2023_birdcage, zhang_2023_diary, sabri2023challenges}; (4) offer instant-reactive and easy-to-use features to protect users during incidents \cite{liao2025_proactive_safety, zheng2023understanding, weerasinghe2025beyond, Chen2025DemocraticModeration, fiani_bigbuddy_2023, blackwell2019harassment}. 

\begin{figure*}[htbp]
     \centering
     \includegraphics[width=\linewidth]{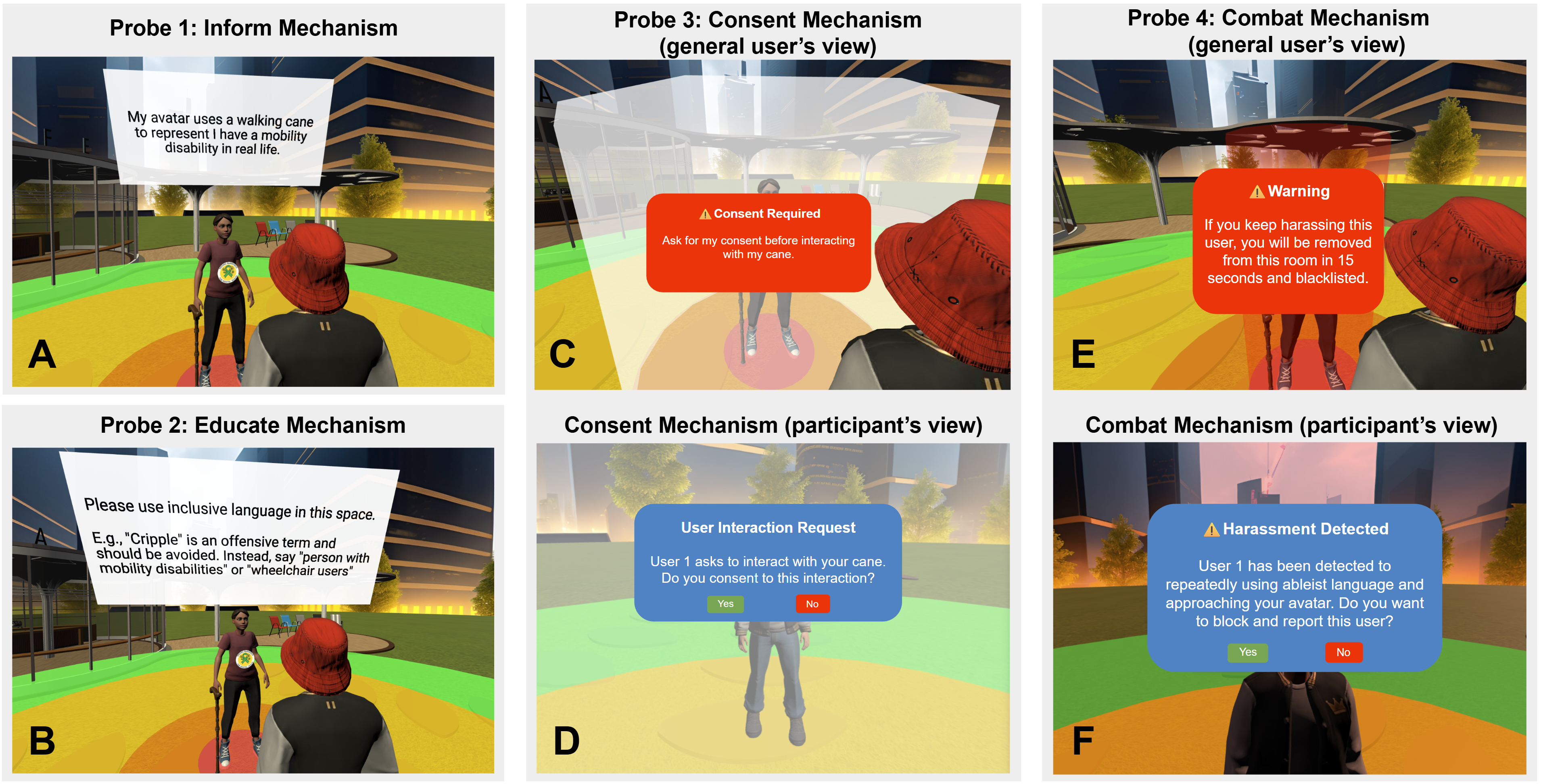}
     \caption{Four Design Probes: (A) Inform Mechanism; (B) Educate Mechanism; (C) Consent Mechanism in general user's view; (D) Consent Mechanism in participant's view; (E) \revise{Defense} Mechanism in general user's view; (F) \revise{Defense} Mechanism in participant's view.} 
     \Description{Figure 2: Six-panel figure showing the study’s VR design probes for proxemics-based protection mechanisms, depicted as in-world pop-up messages between a target avatar with a disability signifier (e.g., cane) and another user. (A) Inform: a speech-bubble style message where the avatar explains the cane represents a real-life mobility disability. (B) Educate: a prompt asking users to use inclusive language, giving an example that “cripple” is offensive and suggesting alternative wording. (C) Consent, general user view: a warning-style overlay stating “Consent Required,” instructing the other user to ask before interacting with the cane. (D) Consent, participant view: a dialog box labeled “User Interaction Request” asking whether the participant consents to the other user interacting with the cane, with Yes/No buttons. (E) Combat, general user view: a warning overlay stating continued harassment will lead to removal/blacklisting. (F) Combat, participant view: a “Harassment Detected” dialog indicating repeated ableist language and approaching behavior, asking whether to block and report the user (Yes/No).}
     \label{fig: probes}
\end{figure*}

Based on these strategies, we designed and implemented four protection mechanism probes that centered on PWD's needs: (1) \textbf{\textit{Inform Mechanism}} that communicates the meaning or purpose of the disability signifier on the avatar to others, preventing potential misunderstandings that lead to inappropriate behaviors. For example, a tooltip floating on top of PWD's avatar saying, \textit{``My avatar uses a cane to represent I am a person with mobility disabilities in real life.''} (Figure \ref{fig: probes}A); (2) \textit{\textbf{Educate Mechanism}} that educates others about inclusive language and sets clear expectations for appropriate behaviors. For instance, a signboard showing: \textit{Please use inclusive language in this space. E.g., ``Cripple'' is an offensive term and should be avoided. Instead, say ``person with mobility disabilities.''} (Figure \ref{fig: probes}B); (3) \textit{\textbf{Consent Mechanism}} asks for PWD's permission before anyone can interact with their disability signifiers (e.g., a virtual cane) to prevent unwanted interactions or unsolicited help. For example, when someone attempted to touch a PWD's cane, a virtual shield appears around PWD's avatar and display a warning to remind others of asking for consent before interacting (Figure \ref{fig: probes}C-D); (4) \textbf{\textit{\revise{Defense} Mechanism}} provides immediate support to help PWD respond to harassment and minimize harm during incidents. For example, if someone repeatedly uses ableist language or attempts to touch the avatar's disability signifier, the system temporarily freezes their avatar. PWD will then receive a prompt to block and report the harasser, while the harasser receives a warning to stop immediately or face removal within 20 seconds (Figure \ref{fig: probes}E-F). 

\subsection{Study Procedure}
We conducted a semi-structured interview with design probes. Studies were held in person at our university campus, with the exception of three participants (P1, P6, P9) who completed the study remotely via Zoom due to their preferences and accommodation needs. Participants who attended remotely were required to have a VR headset for the prototype demonstration. The interview was in English and lasted for about 2 hours with \$25/hour compensation. Each study included three sessions: an initial interview, a scenario-based interview, and a co-design activity. 

\begin{figure*}
     \centering
     \includegraphics[width=\linewidth]{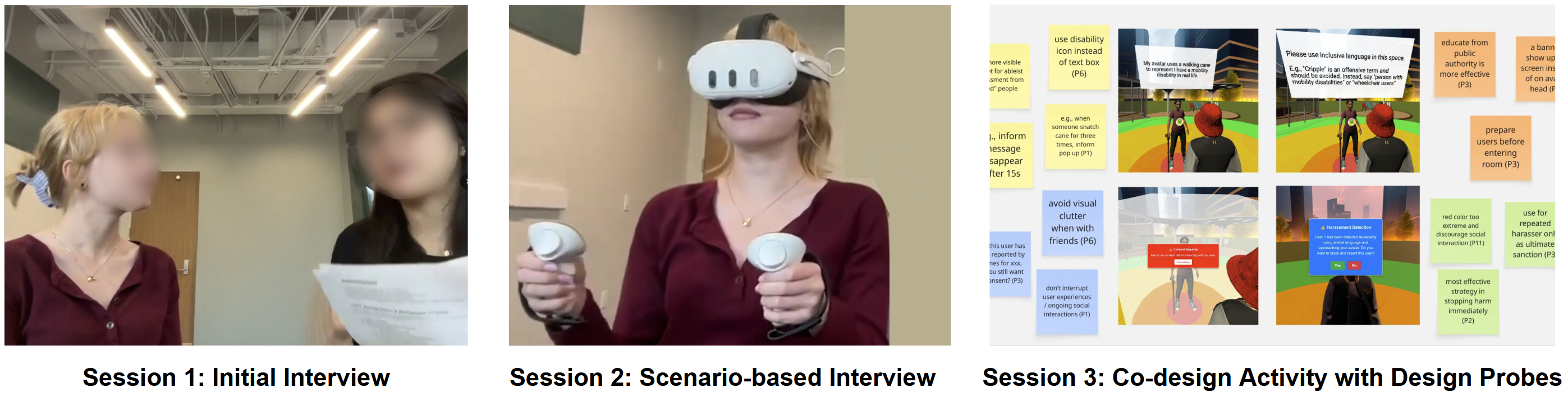}
     \caption{Overview of Study Procedure. We first conducted initial interview with participants to collect their background information and experienced with disability-targeted harassment (left). Then, we handed participants a VR headset and situated them in a virtual environment to view three \revise{common examples of} harassment scenarios; they shared their experiences after each scenario (center). Finally, \revise{we started co-design session with general brainstorming of protection mechanisms, then} presented four design probes to guide discussion and solicit PWD's feedback and needs (right).}
     \Description{Figure 3: Three-panel figure illustrating the study procedure across three sessions. Left (Session 1: Initial Interview): two people sit facing each other in an indoor room; one holds papers, indicating an interview conversation. Middle (Session 2: Scenario-based Interview): a participant wears a VR headset and holds two hand controllers, indicating they are experiencing and responding to VR scenarios. Right (Session 3: Co-design Activity with Design Probes): a collage of virtual-environment screenshots and digital sticky notes, showing participants reviewing example probe interfaces (e.g., inform/educate/consent/combat prompts) and generating design ideas.}
     \label{fig: procedure}
\end{figure*}

\subsubsection{Initial Interview}
We first went through the consent form with participant and obtained their oral consent. We then collected background information, including demographics (i.e., age and gender), self-reported disability conditions, social VR experiences, disability-targeted harassment they have encountered when use social VR and in real life, and their mitigation strategies.

\subsubsection{Scenario-based Interview}
To thoroughly understand the \change{harassment types} and effect of proximity on participants' harassment perception, we contextualized them in a social VR space via a VR headset and \change{illustrated three examples of disability-targeted harassment} 
(Section \ref{sec: harassment scenarios}) at each interpersonal distance (Section \ref{sec: proximity}). 
After viewing each harassment at each distance, we asked participants to rate their sense of safety on a scale of 1 to 5 (1 = very unsafe, 5 = very safe), explained their rating, and described what made them feel safe or unsafe and what harm they sensed. This allowed us to compare how safety perception changed across distances and across harassment types. After experiencing a certain type of harassment at all four distances, we asked participants to reflect across space, identifying at which distance they began to feel uncomfortable or threatened, and when the harassment became intolerable and required action. 

\subsubsection{Co-design Activity with Design Probes}
We then conducted a co-design activity. To thoroughly explore the design space, we began with open brainstorming to capture participants' authentic needs and preferences for protection mechanisms. For each harassment type, we asked participants what support they would prefer at each distance, including both preventive and reactive measures. We encouraged participants to think creatively without being limited by existing technologies or social VR features, focusing on ideal solutions that they wished social VR could provide in the future. We also prompted participants to compare their preferred support across different distances, exploring how their protection needs changed with proximity.

After open brainstorming, we presented the four protection mechanisms as design probes (Section \ref{sec: design probes}) and gathered participants' critiques and suggestions. For each mechanism, we asked participants to discuss whether and how it could be effective in preventing or mitigating each disability-targeted harassment, at which proximity layers this mechanism should be triggered, their preferences on different design aspects (e.g., triggering method, presentation formats) and how they wanted to improve the design. We concluded by asking participants any other desired support they would like to have in social VR, encouraging them to explain their preferences and design considerations throughout the discussion. We listed our full protocol in Appendix \ref{protocol}.

\subsection{Data Analysis}
\revise{We analyzed participant data with a mixed-method approach, combining both quantitative and qualitative analysis.}

\revise{\subsubsection{Quantitative Analysis}
We collected safety ratings at each proximity for each harassment type to obtain a structured measure of perceived safety. We have two within-subjects factors: \textit{Proximity} (public vs. social vs. personal vs. intimate) and \textit{Harassment Types} (verbal vs. physical vs. environmental). Since the ratings were ordinal, we applied the Aligned Rank Transform (ART) ANOVA \cite{wobbrock2011aligned} to evaluate the effects of two factors on perceived safety. We further conducted post-hoc comparisons with Tukey's HSD correction if significant effects were observed. We report effect sizes using partial eta squared ($\eta_p^2$), with 0.01, 0.06, and 0.14 representing the thresholds of small, medium, and large effects \cite{cohen2013statistical}.}

\subsubsection{\revise{Qualitative Analysis}}
Upon participants’ consent, we audio-recorded and transcribed all interviews using an automatic transcription service. Two researchers reviewed the transcripts and manually corrected all transcription errors. We conducted thematic analysis \cite{braun2006theme, terry2017thematic} to identify repetitive patterns and themes in the interview data. First, two researchers open-coded two identical sample transcripts (more than 15\% of the data) independently at the sentence level. We created an initial codebook by discussing and reconciling codes to resolve any differences. \revise{They re-applied the codebook independently to the sample transcripts, reaching reliable agreement with a Cohen’s Kappa of 0.83, then resolved remaining disagreements fully.} 
Next, the lead researcher coded the remaining transcripts independently based on the agreed codebook. 
\revise{When a new code emerged, both researchers discussed to reach agreement, updated the codebook, and revisited prior transcripts for consistency.}
Throughout this process, the lead researcher regularly discussed with the research team to iterate upon and refine the codebook until a complete agreement was reached within the group. 
The final codebook contains 388 codes.

We then derived themes from the codes by combining inductive and deductive approaches \cite{braun2006theme}. Our research has a specific goal of understanding how proximity affects harassment perceptions and protection mechanism preferences. Our high-level theme generation thus focused on these particular aspects \revise{(e.g., proximity layers, harassment types, protection mechanisms),} following the deductive approach. We also adopted the inductive approach and generated themes by grouping relevant codes using axial coding and affinity diagramming. 
\revise{We examined relationships among codes, clustered conceptually similar codes, and iteratively refined them into higher-level themes and sub-themes, yielding six 
themes with 21 sub-themes. We further mitigated interpretation bias by involving a third researcher overseeing the coding process. All authors participated in peer debriefing \cite{creswell_determining_2000} to ensure the themes comprehensively represent the data.} 


\subsection{Ethical Considerations and Positionality} \label{Ethics}
\revise{Since our research focused on harassment perception and protection mechanism design, we paid particular attention to research methodology and ethics to ensure the physical safety and mental health of participants with disabilities.}

\revise{A center piece of our work is proximity, specifically on how the harassment perception and protection references shift as a harasser moves through a 3D social VR space. Such embodied, proximity-graded perception cannot be adequately gauged through 2D-video-probed reflections or retrospective interviews. We thus decided to present three examples of disability-targeted harassment at different proximity layers in a simulated VR environment to elicit participants' harassment perception.} 


\revise{We recognized that exposure to harassment scenarios could place an emotional burden on participants. To minimize potential risks, we chose mild and less graphic examples reflecting common social VR harassment (e.g., using a generic term like ``disabled'' to demonstrate Verbal Harassment rather than explicit ableist slurs).
We also deliberately implemented the following safeguard measures:} (1) Before starting the study, we worked closely with our university's IRB office to prepare the study protocol, ensuring the study objectives and procedures were accurately described;
(2) In the consent phase, we made it clear to our participants that their participation was completely voluntary. Following the best practices of conducting research on online harassment \cite{schafer2023participatory}, we reminded participants that they could skip any questions or withdraw at any time without penalty; 
(3) Before demonstrating the harassment examples, we explicitly told participants that all harassment behaviors were fictional and rendered by the research team for study purposes; \revise{we also verbally described the upcoming scenarios and confirmed their willingness to continue;}
(4) During the study, the research team performed multiple check-ins with participants to ensure they were in a good mental state; 
(5) We also shared resources with participants to cope with online hate before and after the study \cite{ADL_OnlineHateHarassmentSafetyGuide}. 

\change{We value PWD's perceptions and lived experiences and believe that their preferences and insights play a significant role in guiding protection mechanism design. All authors have rich experience conducting research with the disability community, and some research team members identify as people with disabilities themselves, bringing a unique lens in the study design and data interpretation.} 

%% file: sections/tables/demographics.tex
\begin{table*}[t]
\footnotesize
\setlength{\tabcolsep}{4pt}
\renewcommand{\arraystretch}{1.2}
\caption{Participants’ demographics, including ID, age, gender, disability conditions, social VR experiences, harassment experiences in real life (IRL) and social VR, and disability signifiers on avatars.}
\label{tab: demographics}
\begin{tabular}{p{0.48cm}>{\raggedright}p{1cm}p{3.36cm}p{3.36cm}p{4.40cm}p{3.36cm}}
\toprule
\textbf{ID} & \textbf{Age / Gender} & \textbf{Disability Conditions} & \textbf{Social VR Experiences} & \textbf{Harassment Experiences in Social VR \& IRL} & \textbf{Disability Signifiers on Avatar} \\
\midrule
P1  & 32/M   & Severe neuropathy below the left knee; uses cane  & VRChat for 3 years & Encountered ``cripple'' comments and stereotypical disabled avatars in social VR& Cane \\

P2  & 27/F   & Depression, Anxiety, Autism  & VRChat for 3 years  & Stalking, conversation interruptions in social VR & Sunflower symbol for invisible disabilities \\

P3  & 30/NB  & Depression, Anxiety & VRChat for 2 years  & Sarcastic comments of having depression IRL &Sunflower symbol for invisible disabilities \\

P4  & 24/NB  & Connective tissue disorder; ADHD; uses cane & Multiplayer VR game for 2–3 years & Non-consensual interaction with cane and “cripple” comments IRL & Cane, Sunflower symbol for invisible disabilities \\

P5  & 19/M   & ADHD, Autism & Rec Room for 10 hours & Witnessed verbal harassment in social VR; being made fun of and called autistic in school & Orange ribbon for ADHD awareness \\

P6  & 33/NB  & Osteogenesis Imperfecta Type III; uses wheelchair  & VRChat for 4 years; Horizon Worlds and Rec Room for a few hours & Encountered “cripple” comments in social VR & Wheelchair \\

P7  & 22/NB  & ADHD, Autism  & VRChat for 1 month  & N/A & Orange ribbon for ADHD awareness  \\
P8  & 64/F   & Mobility challenges with balance and stride; uses cane   & Immersive Theater app for 1 year &  Being asked to move faster when managing mobility challenges in social VR & Cane \\

P9  & 30/F   & ADHD & Spatial and Engage for 3 years &  Witnessed a group of users trolling in social VR & Orange ribbon for ADHD awareness  \\

P10 & 19/F   & Chronic pain in left leg; uses cane & Custom multiplayer VR app for a few hours & Encountered uncomfortable comments of using cane IRL & Cane \\

P11 & 18/F   & Spina Bifida and Chronic Kidney Disease; Anxiety; Autism  & VRChat for 2 years & Mockery of not understanding social cues and verbal slurs in social VR & Rainbow Infinity symbol for autism \\
\bottomrule
\end{tabular}
\end{table*}

%% file: sections/4-findings.tex
Our findings revealed factors that shaped PWD's harassment perceptions, with closer distances signaling greater threats. Beyond harassment perception, we uncovered PWD's unique needs and preferences for protection mechanisms. 
We detail our findings in the following sections. 


\subsection{How do PWD Perceive Harassment and Harm in Social VR?} \label{harassment perception}
We found that proximity served as a critical signal for PWD to assess harassment in social VR. As a harasser avatar moved closer, PWD increasingly perceived them as threatening and felt directly targeted. Beyond proximity, PWD also used contextual cues such as bystander reactions to evaluate how severe the harassment was. Together, they helped PWD distinguish between ambiguous situations and ableist harassment. 


\subsubsection{Proximity Affects Perceived \revise{Safety}.} \label{safety_rating} 
In social VR, we found that interpersonal distance (i.e., proximity) signaled relationship dynamics similar to real life \cite{hall_1966_hidden}. As a virtual harasser moved closer, all participants felt increasingly \textit{``threaten[ed]''} (P3, P8) and \textit{``targeted''} (P1, P3, P11). 
\revise{Specifically, both \textit{Proximity} ($F_{3,110} = 25.6$, $p < .001$, $\eta_p^2 = .41$) and \textit{Harassment Types} ($F_{2,110} = 4.1$, $p = .020$, $\eta_p^2 = .07$) had a significant effect on perceived safety, with no interaction. We further found that Environmental Harassment elicited significantly higher perceived safety ratings than Verbal Harassment ($t_{110} = 2.72$, $p = .021$). For proximity, perceived safety was significantly higher at public distance than at personal ($t_{110} = 4.49$, $p < .001$) and intimate ($t_{110} = 8.48$, $p < .001$) distances; social distance was perceived to be safer compared to intimate distance ($t_{110} = 5.96$, $p < .001$); and personal distance also yielded higher safety than intimate distance ($t_{110} = 3.99$, $p = .001$). No significant difference was found between public and social distances ($p = .063$) or between social and personal distances ($p = .205$).}
In the following, we unpack \revise{how and why} proximity affects participants' safety perceptions in social VR. 

\textbf{\textit{Proximity Declares Target of Harassment.}}
As the harasser avatar moved closer, all participants were more certain that the harassment were targeted at them. At the Public Distance, eight participants (e.g., P1, P3--6) could not tell who the harasser was targeting, especially when multiple avatars were present in a virtual space. For example, P1 interpreted that a thumbs-down gesture from a harasser could be directing at someone behind him, and P3 described the Verbal Harassment as \textit{``evoking group attention rather than any individuals.''} As P9 explained: \textit{``[At that] long distance, even though [the harasser avatar] is [facing] towards my direction, it doesn't necessarily [target at] me.''} 

However, the target became clearer at the Social Distance, and the harassment felt increasingly direct in the Personal and Intimate Distances (e.g., P2--3, P9). P1 described his perception shift as the harasser avatar moved closer as: \textit{``While [I thought] they were talking to someone else [at the Social Distance], now I definitely feel [that] they are talking to me.''}

\textbf{\textit{Proximity Reveals Ableist Intent.}}
Participants indicated that intent, the perceived purpose behind harassing behaviors \cite{blackwell2019harassment}, was clarified by proximity. With the harasser avatar at Public or Social Distance, five participants (P2--5, P9) found their behavior difficult to perceive due to far distances, and they were confused about what the harasser was doing (P5). However, as the harasser avatar breached the Social Distance, such confusion and ambiguity started dissolving into a clear perception of ableist harassment, according to all participants. For example, P4 noted that invading personal space made the mocking intent unmistakable: \textit{``To have gotten this close to single me out then change to this [mocking] avatar, there is no ambiguity that this is a joke [and is] intended to harass [me].''}


\textbf{\textit{Personal Distance as the Bottom Line.}}
Notably, nine participants (P1, P3--7, P9--11) identified the Personal Distance as their bottom line, denoting the threshold where harassment became directly targeted, clearly harmful-intended, and no longer tolerable. P4 highlighted the unique harm perceived in Personal Distance by comparing it with further zones: \textit{``The rudest behaviors and the most intense attacks all happen in the Personal Distance. [However,] at the Social space [and further], the delineation between joke, regular interaction, and harassment become blurrier.''} Moreover, Physical Harassment in VR becomes more intense and threatening at the Personal Distance. All four participants (P1, P4, P8, P10) who had a virtual cane on avatar felt that \textit{``the cane is actually snatch-able''} (P1) at the Personal Distance. P4 and P8 particularly feared that the harasser would snatch the cane, throw it back, and hit them with it. 

To prevent further harm, three participants (P5, P9, P11) felt the urge to actively respond once the harasser invaded the Personal Distance. 
As P11 described: \textit{``They're obviously not going to stop if they're getting that close to me, [and] they think what they're doing is okay. In that sense, [I] don't know how far they're going to take it, [and] they may start [more] harassing [behaviors].''}



\textbf{\textit{Virtual Closeness Raises Discomfort even without Harassment.}} 
Interestingly, six participants (P1--4, P10--11) highlighted that being virtually close to some unknown avatar itself made them feel uncomfortable, regardless of the avatar's behaviors. For example, when the harasser avatar was in the Intimate Distance, P1 felt \textit{``uncomfortable about them being that close [rather] than what they're saying.''} P4 further detailed their expected relationship in the Intimate Distance: \textit{``There is an implication of trust when you are so physically close to somebody else. Especially in VR [where] we have much more real perception of proximity than you get from 2D environments, it's all-encompassing [and] feels much more physical. I can't imagine that this person wants anything other than love [by being that close to me].''} Moreover, P1 and P4 noted that such closeness can feel even more concerning in social VR than in real life, because avatars can \textit{``clip inside to each other''} (P4) instead of being forced away by physical bodies.

Our findings extended Hall's Proxemics Theory and suggested that physical-spatial norms can apply to virtual environments, where such a close distance is reserved for trusted individuals (e.g., partners, families, and friends), and a stranger appearing there was an intrusive and increased level of discomfort \cite{hall_1966_hidden}.




\subsubsection{Leverage Contextual Cues to Evaluate Harm Severity.} 
While proximity explicitly signaled the target of a harassment behavior (i.e., myself or others) and intent (i.e., ableist or not), contextual cues helped participants gauge the severity of that behavior. We identified two contextual cues that affected the perceived severity.

\textbf{\textit{Amount of Invested Efforts.}}
Five participants (P2--P4, P9, P11) indicated that the amount of effort and time invested in a harmful behavior revealed how malicious it was. Specifically, they perceived Environmental Harassment as particularly harmful, because \revise{they immediately recognized the ableist intent and found} creating or obtaining a mocking avatar required more effort than uttering an ableist term or performing a demeaning gesture. For example, P11 highlighted the extra effort needed for Environmental Harassment compared to Physical Harassment: \textit{``Pointing and thumbing down doesn't feel terrible, because anyone can just [do it]. But making an avatar [to mock] someone disabled feels more serious. [Because the harasser] took their time to change the avatar to mock me, and they put effort into [being] disrespectful.''} 
P3 further emphasized that the Environmental Harassment had broader harmful impact: \textit{``This is out of interpersonal [scope], because there is something bad in the environment. The whole room is rotten, and everyone is endangered. As long as this [offensive avatar] is visible in the virtual space, I would feel like `oh, there is a bomb in this room.' ''} 

\revise{We noted that Environmental Harassment could be uniquely difficult to recognize and safeguard. We found five participants (P1, P3, P6, P7, P10) who needed to approach the mocking avatar before they could see its details, and some still did not fully understand its mocking meaning because they were unfamiliar with the meme. This delayed or incomplete recognition may help explain the higher perceived safety ratings for Environmental Harassment (Section~\ref{safety_rating}), as participants may have rated the scenario as safer before fully recognizing it.}

\textbf{\textit{Bystander Reactions.}} 
Beyond the harasser's behavior, two participants (P3, P5) leveraged bystanders' reactions to gauge the overall safety and \textit{``vibe''} (P3) of a virtual space. In spaces they perceived as safe, participants expected bystanders to actively intervene when harassment occurred, rather than ignore it or join the harasser. For instance, P3 described how they read the room during Verbal Harassment:
\textit{``I will largely rely on bystanders [to know] the norm and tolerance [of this space]. [Especially] when someone verbalizes their harmful intent out, they must want to grab people's attention. If nobody reacts to them, that [means at least] this room is safe. Ideally, people will start gathering at the harasser [and call out like,] `What's wrong with you?' [But] if people [start] echoing back with [the harasser], that [means this place is definitely] dangerous.''}

\subsection{Managing Harassment with Optimism and Resilience} \label{sec: resilience}
We found that some of our participants (e.g., P1, P6, P11) often responded to harassment with a positive, resilient mindset in VR. Given the prevalence of harassment in social VR, they adapted by showing high tolerance and attributing incidents to ignorance or joking rather than ableist intent. They also believed that the harasser's behavior were correctable through appropriate education rather than sanction. In the following, we detail how this resilience shaped their interpretations of harm and their preferences for protection mechanisms.

\textbf{\textit{Normalizing Ableist Hate in Social VR.}}
Four participants (P1--2, P4, P6) became desensitized to recurring harassment in VR due the lack of social norms in this new social space \cite{zhang_2023_diary}. For example, P1 was not surprised to see offensive avatars after using VRChat for three years. Instead, he thought it was common and not invasive enough for him to respond to it. Frequent exposure to harmful behaviors also made participants develop a mindset to normalize disability-targeted harassment. Similarly, P6 was used to avatar-based mockery, and they coped by laughing to avoid emotional harm: \textit{``Somebody used an avatar meant to mock me is nothing new, and I'm used to [it]. If what [they did] makes me laugh, I will [just] laugh. Sometimes that's better than getting really offended and getting back at them, or giving them [a reaction] they want.''}


\textbf{\textit{VR Harassment is Easier to Handle than Real-World.}}
Despite the high frequency and embodied nature of harassment in social VR \cite{blackwell2019harassment}, four participants (P1--2, P4, P6) viewed them as lower-stakes and more controllable than in real-life encounters, as virtual harassment doesn't have intrinsic physical harm. For example, P6 would only feel dangerous if someone tried to push their wheelchair in real life but not in VR: \textit{``If [they push my avatar] in real life, absolutely [I'd feel unsafe]. In VR, however, I would ask them what they are doing, [because] it's a [virtual] place. I feel awkward if somebody trying to push my wheelchair, which is impossible.''}

Although some harassment caused emotional harms such as anxiety (P1, P4, P6--7), participants found them lasting \textit{``no longer than 3 minutes''} (P1) and easy to deal with by simply muting and blocking the harasser (P4, P6). At the worst-case scenario, P4 and P7 would take off their VR headset to avoid further harm, whereas they would urgently seek bystander interventions in real-life encounters. 

\textbf{\textit{Presuming Non-ableist Intention.}}
Three participants (P1, P4, P6) attributed the harassment behaviors to joking or lack of knowledge of disability, rather than an ableist intent. Recognizing low public awareness of disability, P1 and P6 believed that seemingly harmful behaviors often reflect ignorance. As P1 emphasized: \textit{``I truly believe there is a large amount of ignorance [about people with disabilities] that exists inside the VR space and real life, unfortunately. People just need to be told [about] what you should be saying.''} 

Beyond ignorance, P4 and P6 also framed some incidents as online joking rather than harassment. For instance, P4's initial reaction to Environmental Harassment was: \textit{``this could be a within-community joke [done by] another disabled person''} meant to be self-deprecating rather than mockery.

Interestingly, P6 resonated with the Environmental Harassment's meme avatar, relating it to their own experience of being misread as a ``troll'' when using an avatar with a virtual wheelchair because disability-representing avatars were rare to see in social VR \cite{zhang_2023_diary}. P6 explained: \textit{``I have been in a [virtual] world where people saw my wheelchair and thought that I was a troll. Because [people] don't see avatars [with disability representation] in any virtual setting very often. It is not always immediately apparent whether or not [an avatar with a disability signifier] is a serious representation of somebody or a meme. I get where [the trolling impression is] coming from.''}


\textbf{\textit{Giving Chances for Behavior Correction.}}
Believing that most harassment behaviors were unintentional, six participants (P1, P3, P5--6, P10--11) preferred a buffer period before taking action against the harassers, protecting themselves from immediate harm while still giving the harasser time to correct their behaviors. For example, P6 would first initiate a conversation with the harasser to observe for behavior changes: \textit{``I'll try to talk to them and see how they respond to me. If I make a comment like, `Hey, I'm actually in a wheelchair in real life! It's cool you found an avatar with a wheelchair.' Whether or not it's silly or meant to be a meme or whatever, I will try to make it a positive experience first. [But] if it continues to be negative and harassment, that's when things would change.''}

As a result, four participants (P1, P3, P6, P11) preferred a stepwise approach, warning harassers of negative consequences before applying full sanctions. 
They believed that harassers needed guidance on what to do instead of immediate sanctions. As P3 explained: \textit{``
If those harassers know the negative consequence, such as being blocked or or expelled from this room, they will behave themselves. [But] the drawback is [that] they do not know how to behave better. [So] I would like to give them the warning first.... I'd warn them that `The avatar you are using is not acceptable at this place. Please [change it] within 30 seconds.' [For Verbal Harassment,] it should tell them what terms they violated in this room and what they can say to avoid being so offensive in the future.''} 






\subsection{Pursuing Safety through Risk Avoidance and Systematic Effort} \label{sec: avoidance}
In contrast to the resilient optimism, six participants (P2--4, P7--9) more cautiously managed personal safety by proactively anticipating and avoiding risks. Their caution stemmed from years of facing disability-targeted harassment in real life, from everyday microaggressions to ableist discrimination, which made them vigilant about potential threats when entering social VR (P2--4, P9). 
The anonymity of social VR and thus the more explicit expression of ableism \cite{zhang_2023_diary} further amplified their fears (P3). 

Given the heightened vulnerability, participants prioritized protective strategies over confrontation. Rather than relying on individual efforts to correct harmful behaviors, they believed systemic interventions that establish inclusive norms were the fundamental and effective solution. We unpack  participants' preferences below. 

\textbf{\textit{Minimizing Attention Drawn to Disability.}}
Three participants (P2--4) hesitated to disclose disabilities and avoided using protection mechanisms that may signal their disabilities (e.g., using the Inform or Educate Mechanism), 
which made them a more visible target for harassment. For example, 
P3 was concerned that such mechanism would be misused: \textit{``It's definitely a backfire. Because at least from my experience using VRChat, trolls are everywhere. They will troll around your Inform statement.''} As a result, participants preferred mechanisms without explicitly mentioning their disabilities, so that they could minimize unwanted attention and avoid further harassment. 



\textbf{\textit{Preemptive Distancing to Avoid Escalation.}}
We observed that five participants (P2--4, P6, P11) were alert when they first noticed the harasser and chose to proactively distance themselves away from the situation before any escalation might happen. Participants adopted a range of strategies to disengage themselves, such as physically stepping away from the harasser (P3), teleporting to a different virtual world (P4, P6), and/or completely exiting VR by taking off the headset (P4). After using social VR for a few years, P6 found that avoiding harassers actually worked more effectively than responding back. They explained: \textit{``My experience with trolls [is that] the more you react to a negative situation, [the more it] escalates the situation. So the best way is to ignore them.''} 

In addition, three participants (P1, P4, P7) used safety features provided by platforms (e.g., mute, block) to save mental energy in dealing with harassers. As P4 indicated: \textit{``I'd rather just to block this person and not have [to deal with it]. I just have no incentive to give them the opportunity to change their behavior when I can just fully disengage. I can just block them, [and] that's always what I prefer to do instead of striking them, especially to virtual strangers.''}

\textbf{\textit{Fearing Retaliation from Confrontation.}}
Although four participants (P1--3, P10) desired to correct harassers' behavior and teach them about disability, they often held back because they felt that \textit{``people don't want their views to be challenged''} (P3) and worried that speaking up would trigger retaliation. For example, P2 was concerned that educating a virtual stranger could backfire: \textit{``I would like to verbally correct [the harasser], such as explaining what they did wrong and why it's considered as offensive. [But doing that to a] random person in a public world that I didn't know at all may [lead to] more chance for backlash and [...more] messing up with people.''} P10 echoed this concern, imagining how a harasser might interpret a call-out: \textit{``[They might think] `oh this person wants us all to stop using this language. So I'll go target [at] them now.' ''}



\textbf{\textit{Establishing Inclusive Norms from Authority.}}
To fundamentally improve safety and make PWD feel comfortable disclosing disabilities, six participants (P2--3, P5--6, P10--11) emphasized the need for inclusive social norms. They described these norms as clear expectations for respectful language and behavior (P2), standards for what counts as disability-targeted harassment (P11), and shared accountability for bystander interventions (P3, P5).

Four participants (P1, P3, P5, P11) particularly highlighted that inclusive norms should be conveyed by an authoritative source instead of an individual PWD, because it felt more credible, legitimate, and hard to ignore. For example, P3 argued that the Educate Mechanism should be a system-driven action instead of a personal action. They explained: \textit{``Being taught by a random person is not as effective as the system. [People will] question the reliability or the credibility of the source.''} P11 also found Educate Mechanism with an authoritative design is more effective, as she envisioned: \textit{``It sets the standard. I'm imagining it shows as a banner when people join the world. I think showing it to everyone, rather than making it an individual thing, will make it a safer space for everyone involved.''} P1 and P5 further emphasized that norm-setting mechanisms need to be attention-catching, such as a large banner at the entrance of a virtual space (P5) or as an audio announcement to everyone before entering the world (P1). 

\subsection{Responses and Protection Mechanisms as a Medium for Self-Presentation} \label{sec: self-presentation}
\revise{Our findings highlighted that many participants} cared deeply about how they appeared to others when responding to harassment. Years of experiencing disability-related stigma (e.g., being seen as overly sensitive, vulnerable, or preachy \cite{Manno2024DisabilityDisclosure}) 
motivated them to strategically craft their responses to harassment behaviors to challenge these stereotypes and establish desired social identities. \revise{For example, few participants (e.g., P2, P4, P9) avoided responding to harassment to not make a scene; while others (e.g., P1, P5, P10)} leveraged their responses as opportunities to \revise{project more assertive social image. In this session, we outlined how participants} present themselves as a playful, friendly, or powerful figure depending on contexts. 
These social image concerns directly shaped their protection mechanism preferences, favoring mechanisms that could protect them without reinforcing the stigma they worked to resist. 

\textbf{\textit{Playful Figure to Look Chill with Disabilities.}}
When encountering harassment, two participants (P1, P6) responded with humor or joked back to present themselves as relaxed or calm. For example, when the harasser tried to snatch his virtual cane, P1 replied causally: \textit{``Oh you wanted to try that out? But you can't have it.''} Similarly, P6 would \textit{``tease them back and laugh it off''}  when harassers tried mocking them with meme avatars.

P1 and P6 intentionality avoided being perceived as overly serious with their disability or coming across like a \textit{``preacher''} (P1), a common assumption others made about PWD who spoke up about disability. P1 explained in detail: \textit{``I don't want [people] feel like I'm being uptight or overly stuck up about [my disability]. It should just be casual. [I don't want people to think of me like], `Are you a textbook? Why are you talking to me like that?' That's not normal.''} This approach was especially helpful in playful social VR contexts, where P1 and P6 wanted to fit the tone of the room while still signaling comfort with their disabilities. 

 
\textbf{\textit{Friendly Figure to Facilitate Positive Social Vibes.}}
Three participants (P3, P6, P9) preferred to respond in a friendly, upbeat tone to sustain positive vibes and reduce the chance of escalation, especially when intent was ambiguous. For example, P3 preferred using an informing message that was \textit{``lively and cute''} to explain the meaning of the disability signifier on their avatar, framing it as encouragement rather than correction. They gave an example of how they may communicate their avatar design to a potential harasser: \textit{``[I would say]: `The ribbons and sunflowers [on my avatar] are for soothing people [with] mental health struggles. They're doing a good job, and you are also doing a good job [by providing] your support.' ''} Similarly, P6 tried to avoid being defensive when encountered the meme avatar in a wheelchair; instead, they started with a friendly comment to invite meaningful conversations: \textit{``I would be friendly, because it isn't necessarily always harassment. My first reaction wouldn't be defensive. It would actually be, `Oh, cool, you got a chair like me. That's awesome!' ... I would probably make a more positive comment at first to gauge what kind of interaction I'm gonna have with that person.''}

\textbf{\textit{Powerful Figure in Defending Disability Community.}}
However, when a harasser crossed the bottom line, \revise{three participants (P3, P5, P10) chose to correct the misconceptions and even} fight back to present themselves as a powerful defender of the disability community. Interestingly, P3 felt especially empowered to do so at the Personal Distance for two reasons: First, being close created more equal conditions for confrontation because both parties could clearly see and hear each other: \textit{``Within [the] personal space, [we can] clearly hear and see [each other], so it is a proximity that is not safe for me, nor safe for them.''} P3 framed this as a mutual vulnerability that reduced the harasser’s advantage. More importantly, P3 noted that they would be less likely to fight back in real life due to physical vulnerability and limited ability to defend themselves, whereas VR removed this constraint. After enduring stigma about mental health for years, P3 described VR as a space where they could finally push back and feel proud to stand up for the mental health community.

\subsection{Desiring Intelligent, Socially-Aware Protection Mechanisms} \label{sec: desire}


Across the co-design session, participants converged on a shared vision: protection mechanisms should not only react to proximity, but should understand social context, respect their desired social image, and support rather than disrupt social interaction. We detail four qualities participants considered essential below.

\textbf{\textit{Be Proximity-Aware to Prevent Escalation.}} 
Eight participants (P2--6, P9--11) emphasized that protection mechanisms should be proximity-aware, providing support before a harasser crosses their boundary and causes further harm. 
Participants detailed their needs and preferences across distances. At greater distances such as the Public and Social Distance, three participants (P3, P6, P11) preferred lightweight monitoring of potentially harmful behavior. They explained that there was no immediate threat at these far away distances; thus, monitoring simply helped them stay alert without being intrusive. P2 and P9 found the Public Distance particularly suitable for the Educate Mechanism, which established behavioral expectations in highly visible ways so that everyone in the space can see them. At the Social Distance, two participants (P3, P5) mentioned the value of third-party intervention support, such as involving bystanders or room moderators, to help mediate situations before they escalate. However, once interactions reached the Personal Distance, participants (P2--6) wanted more decisive and stronger support, such as automatically blocking and muting the harasser. They explained that at this proximity, it becomes harder to simply walk away, and the risk of unpredictable Physical Harassment increases significantly (P10, P11).

\textbf{\textit{Be Intent-Aware to Avoid Over-Punishment.}}
Four participants (P1, P3, P10--11) worried that protection mechanisms could be accidentally triggered by people without harmful intent, leading to unnecessary interruption or over-punishment. For example, P3 noted that a Consent Mechanism might be triggered when someone was simply wandering past through a space, rather than intentionally approaching them; P1 and P4 described cases where friends with disabilities might use terms like ``cripple'' as an in-community joke. 
In these situations, participants highlighted that the activation of protection mechanisms should be cautious and intent-aware, so that mechanisms do not discourage positive interactions.

Therefore, participants (P1--4, P6, P11) suggested that protection mechanisms should leverage the rich signals from social contexts to infer user intent. For example, P1 suggested employing the Consent Mechanism if a person tried to touch their cane three times, which confirmed their interest for interaction rather than accidentally running into their avatar. P2 and P3 proposed using a user's historical behavior pattern as a signal for assessing intent. They envisioned a safety credit system for each user, and protection mechanisms would trigger more often for users with low scores. As P3 indicated: \textit{``If the requester is generally a nice person, I think the consent is unnecessary. It's just a normal conversation to reduce the barriers for both them and me. And I don't want to approve everyone. But [consent is required] for those very dangerous [users], and I can set that threshold [based on their safety credit history].''} 



\textbf{\textit{Be Context-Aware to Respond in Rapidly Escalating Situations.}}
Four participants (P4, P6--7, P11) emphasized that protection mechanisms must respond quickly in situations where harassment escalates rapidly, as manually activating safety features under stress can be cognitively overwhelming. P4 envisioned an AI-powered automation to reduce this burden: \textit{``Pulling up and blocking something is more difficult than just [responding to] `Hey, it looks like this person is not treating you very well. Do you want to block them?' AI could be useful for [such] emergent situations.''} Similarly, P6 wanted proactive support triggered by proximity violations, noting that threatening encounters often occurred when harassers invaded their personal space: \textit{``Every instance where I felt threatened or encroached on is usually when they're really close. Maybe just having a little pop-up [when they are close], that's like, `Hey, do you want to record this interaction?' That being semi-automatic [means] the user still has a little bit of control, where it's not just happening [completely] automatically.''} Both participants preferred protection mechanisms that automatically predict escalation and offer proactive assistance while preserving user agency in the final decision.

\textbf{\textit{Be Pro-Social.}}
Six participants (P1--4, P6, P11) noted that protection mechanisms could visually crowd and disrupt others’ experiences in the shared VR space. Although these mechanisms were intended for safeguard, participants tried to minimize the burden their protections placed on others.

As a result, participants preferred subtle protection mechanism designs. For example, to explain the meaning of their virtual wheelc-hair in the Educate Mechanism, P6 preferred an interactive disability icon over a big text box. As P6 envisioned: \textit{``Instead of it being a big text box, having an icon [of] the universal disability insignia [next to the wheelchair on my avatar] would be better. [People can] tap on it, and then it shows the text box. That way, it's not big and gaudy over somebody's head all the time, which draws a lot of attention and could invite people to poke fun.''} P1 and P4 suggested that protection mechanism design should only show up for limited time periods (e.g., for a 10-15 second window to allow others to read) then disappear to reduce visual clutter. Finally, P3 and P11 recommended using a less intrusive color scheme (e.g., avoiding red) so that safety cues would not discourage social interactions. 

%% file: sections/5-discussion.tex
This paper contributes the first work in understanding PWD's values, needs, and preferences for protection mechanisms in social VR. Inspired by Hall's Proxemics Theory \cite{hall_1966_hidden}, our results revealed that proximity served as a rich socio-spatial signal that PWD used to assess harassment target, intent, and severity (Section \ref{harassment perception}). Through co-design with four design probes, our findings surfaced PWD's diverse preferences for protection mechanisms, ranging from those who favored mild preventions that preserved a sociable self-image to those who preferred systemic, depersonalized safeguards that minimized their visibility as a target (Section \ref{sec: resilience} \& \ref{sec: avoidance}). Importantly, we found that protection mechanisms serve as both safety tools and medium for self-presentation, through which PWD managed how others perceived them while combating (or tolerating) harassment (Section \ref{sec: self-presentation}). These findings highlighted a core design requirement for disability-centered protection mechanisms in social VR, ensuring their personal safety while preserving desired social image.
In this section, we discuss the uniqueness of PWD's needs for protection mechanisms as well as the design considerations to facilitate a safe social VR environment.

\subsection{Rethinking Proximity in Harmful Social VR Interactions}
Our research makes a theoretical contribution by understanding harassment and protection mechanisms through a social lens. We extend Hall's Proxemics Theory \cite{hall_1966_hidden} into an important yet underexplored context---harmful interaction between PWD and malicious users. We consider harassment itself as a form of social interaction, but with a hostile intent instead of a cooperative one. We unpack these dynamics below. 

Hall's Proxemics Theory \cite{hall_1966_hidden} correlated physical proximity with types of social interactions that typically happen within that distance, such as the distance for intimate interactions versus conversations. Prior social VR research has inherited this framing by using proximity to support friendly and collaborative interactions \cite{williamson2021proxemics, ji2022vrbubble}. 
In our work, however, the same proximity structure governed hostile encounters. As a harasser's avatar moved from public toward personal distance, PWD read proximity as a progressive disclosure of three things: who was being targeted (from ambiguous to directly personal), what was the intent of the behavior (from unclear to recognizably ableist), and how urgently they needed to respond (from monitoring to demanding immediate action). 

In this light, proximity is not only a guide for designing sociable interaction but also a framework for interpreting harassment and for guiding protection mechanism design. 
Unlike prior work that treated proximity as a static threshold for setting a personal boundary \cite{zheng2023understanding, sharma_2022_personalboundary}, we position proximity as a spatial continuum that sheds light on which protection mechanism is suitable at which distance to better prevent harm from happening. Our study captured a range of potential designs centered on proximity (Section \ref{sec: desire}). For example, Public Distance fits best for proactive norm-setting mechanisms (e.g., community guidelines, code of conduct) due to its visibility, establishing behavioral expectations for a broad audience before personal interactions may happen; Social Distance, however, needs just-in-time intervention that stops the harmful action when it is about to occur. This proximity-driven design framework moved the protection design a step further toward preventive intervention before harassment escalates. 


\subsection{Protection Mechanisms as a Medium of Self-presentation}
Our work introduces a new perspective: protection mechanisms are not merely safety tools but also social gestures that calibrate to how PWD want to be perceived by others. Prior research has found PWD often manage how openly they disclose their disability, since visibility may invite stigma that hurts their social image \cite{goffman1963stigma, joachim2000uncertainty, nario2019hostile}. Our findings echo and expand existing literature by demonstrating that, even in embodied social VR settings where harassment can escalate within seconds, PWD still weighed how the use of protection mechanisms may impact their social image before deploying them. We observed many cases where participants sacrificed personal safety to better curate a positive social image (Section \ref{sec: resilience}). For example, rather than seeking protection mechanisms when they encounter harassment, PWD strategically downplayed incidents by attributing hostility to ignorance instead of ableist intent to minimize conflict. They also responded to mockery with humor to appear unbothered and avoid being perceived as a preacher about disability. Echoing prior literature \cite{shakespeare1999joking}, this coping strategy serves as a resistance practice that allows PWD to reclaim the interaction on their own terms. Even when engaging with protection mechanisms, PWD preferred them to be subtle and soft, avoiding drawing attention to disabilities or creating a scene that might lead others to perceive them as oversensitive or punitive.

PWD's prioritization of social image reflects the structural costs attached to visible disability protection \cite{corrigan2002paradox, narioredmond2013redefining}. Disability studies literature has long documented the stereotypes that attach to PWD in everyday life (e.g., as someone to be pitied or cared for \cite{barnes1992disabling}). These stereotypes create a social environment where rejecting patronizing help is penalized \cite{wang2015independent} and visible accommodation needs are framed as markers of dependence \cite{nario2019hostile, sanders2006overprotection}. 
The weight of these externally imposed identities may push PWD to treat visible or confrontational protection mechanisms as a social cost that confirms the stereotypes they work to resist. Taken together, we highlight that protection mechanisms serve as a medium of PWD's self-presentation. Designing for this dual function requires mechanisms that effectively mitigate harm while preserving PWD’s desired social image, and should do so without imposing additional emotional strain.


\subsection{Towards Protection Mechanism that Preserves Social Image}
Based on PWD's needs and preferences, we derive design implications to inspire pro-social protection mechanisms that better ensure safety while preserving their ideal social image. 


\textbf{\textit{Detecting Intent instead of Action.}}
Protection mechanisms should detect hostile intent rather than just actions. With the recent advancement of multi-modal AI \cite{potla2025ai}, existing systems can flag harmful actions from voice \cite{xu2024safe}, gestures \cite{wang2024hardenvr}, and avatar visuals \cite{lee2025harassguard}, yet a single flagged action does not reliably indicate harassment. Our findings suggest that PWD read intent not from isolated behaviors but from behavioral patterns (e.g., a snatching gesture happened three times, a user with multiple report record of harassing others). In this way, the system should be better at distinguishing genuine hostility from one-time misbehavior, avoiding over-punishing people and resisting the overly sensitive stigma attached to PWD. Future protection mechanisms should thus aggregate behavioral signals across time and across users to more accurately identify ableist intent. 

\textbf{\textit{Handling Conflicts Privately.}}
PWD preferred resolving conflicts privately to avoid disturbing others' social experiences. Our participants often hesitated to engage with visibly confrontational protection mechanisms (e.g., \revise{Defense} Mechanism). They are concerned that doing so would draw additional attention to their disability and potentially damage their social standing as well as future interactions within the community. Therefore, future protection mechanisms should consider to be discrete and privacy-preserving. One way to do this is to leverage the spatial affordance of social VR to contain interventions away from public view. For example, when a potential hostile behavior is detected, the system could create a temporary Private Bubble around the two individuals, rendering their conversation inaudible and invisible to others so that the situation can be addressed without an audience. For PWD who prefer not to engage at all, platforms could offer graceful exit mechanisms that smoothly transition them to another area or session, preserving their agency without a conspicuous departure.

\textit{\textbf{Offloading Social Cost to the System.}} Existing platforms rely on user-initiated actions such as blocking or vote-kicking \cite{blackwell2019harassment, freeman2022disturbing, weerasinghe2025beyond, Chen2025DemocraticModeration}, which put PWD at risk of retaliation according to our participants. To mitigate such concerns, future protection mechanisms should not attribute their activation to individual users, but frame protections as system-level actions, so that the social cost of intervening is carried by the platform rather than individual PWD. This framing could be particularly important when the intervention is corrective or confrontational, where the more forceful the action, the more social cost it may bear \cite{czopp2006standing}. Platform-mediated mechanisms also carry greater authority and legitimacy, making users more likely to accept and comply with the behavioral expectations.

\textbf{\textit{Deterring Harm through Social Accountability.}}
Protection mechanisms should deter misbehavior by making harassers visibly accountable for their actions. This is especially important in social VR, where harassers can harass and disappear right away \cite{blackwell2019harassment, zhang_2023_diary}, facing no consequences for targeting PWD. To prevent hit-and-run harassment incidents, platforms could leverage proximity as a trigger for accountability cues. For example, when a flagged user enters the Social Distance, which is the ``bottom line'' according to PWD, the system could prompt the PWD to begin recording the subsequent interaction. The recording preserves evidence if the situation escalates, and the act of recording also signals to the harasser that their behavior is being documented. This shifts the social cost of misconduct onto the harasser and gives PWD a buffer to de-escalate without direct confrontation.

\textbf{\textit{Normalizing Disability Representation in Social VR.}} 
Beyond correcting individual behaviors, PWD believed that fostering inclusive norms would be a more fundamental, long-term solution. While norms take a long time to evolve and establish \cite{kiesler2012regulating}, protection mechanisms can start by increasing the visibility and normalizing disability presence in social VR. For example, platforms could populate virtual spaces with Non-Player Characters (NPCs) that carry representative disability signifiers like virtual wheelchairs, white canes, or guide dogs, and assign these NPCs to various roles such as greeters, guides, or ambient participants. For PWD, seeing avatars with disability representations throughout a space also signals that their presence is welcome and expected, thus reducing the sense of isolation. Platforms could further design embodied AI moderators with disability representations, positioning PWD in roles of authority and community leadership rather than as subjects of protection, directly countering the stereotypes of vulnerability and dependence that PWD worked to resist \cite{goffman1963stigma, daley2018stigma}.

\subsection{Limitations and Future Work}
Our research has limitations. First, we examined perceptions of harassment and preferences for protection mechanisms through fictional scenarios we scripted and rendered in a controlled social VR scene. We recognize that the controlled scenarios lack the messiness of live interactions, falling short in capturing evolving dynamics and social stakes. \revise{For example, the use of genderless and ethnically ambiguous avatars overlooks how intersectional identities shape harassment perception and protection preferences.} Future work should consider to conduct a longitudinal study in active social VR communities \cite{zhang_2023_diary}, which could yield richer evidence through observations and in-the-wild data. Moreover, while we followed a common practice to visualize the proximity, we acknowledge that use of color (i.e., the green-to-red gradient from public-to-intimate distances) may have primed participants' safety perceptions, associating closer distances with danger. Future work should explore more neutral visualizations such as numbered layers. Lastly, we realize that our sample is limited in size and diversity, which did not capture the full range of disability experiences. Future work may consider involving larger and more diverse PWD cohorts, cross-cultural and neurodivergent samples, and novices alongside experienced users to further enrich the study data and foster greater generalizability.

%% file: sections/6-conclusion.tex
In this paper, we conducted a co-design study with 11 PWD to investigate their needs and preferences for protection mechanisms in social VR. Grounded in Hall’s Proxemics Theory, our findings revealed that proximity operates as a rich socio-spatial signal through which PWD read the target, intent, and severity of disability-targeted harassment. Beyond proximity, participants’ responses were shaped by divergent social values, ranging from managing harassment with optimism and resilience to seeking systemic, depersonalized safeguards. Importantly, we found that protection mechanisms functioned not only as safety tools but also as a medium of self-presentation, through which PWD negotiated a persistent tradeoff between personal safety and their desired social image. Building on these insights, we proposed design recommendations that decouple protection from involuntary disability disclosure, offload the social cost of intervention to the platform, and normalize disability representation in social VR, charting a path toward protection mechanisms that center on PWD's needs. 

%% file: sections/7-appendix.tex
\subsection{Disclosure of AI Usage}
AI tools such as Claude and ChatGPT were used to assist writing (e.g., correcting grammar, finding a more accurate word) in this manuscript. All content has been thoroughly edited and verified by the research team for accuracy. 

\subsection{Study Protocol}  \label{protocol}
\textbf{Part I: Demographics \& Background}
\begin{enumerate}
    \item What's your age?
    \item How would you identify your gender?
    \item What's your ethnicity?
    \item What disabilities do you have?
        \begin{itemize}
            \item How long have you had the disabilities?
            \item How do they affect your daily life?
            \item Do you use any assistive technology in daily life?
        \end{itemize}
    \item What VR devices have you used before?
    \item What social VR applications have you used?
        \begin{itemize}
            \item What's your most commonly used social VR application(s)?
            \item How long have you been using them?
            \item Who do you usually socialize with on this platform?
        \end{itemize}
    \item Have you ever disclosed your disability identities while using social VR?
        \begin{itemize}
            \item If so, how did you disclose your disability identities?
        \end{itemize}
    \item Can you recall the most memorable experience of disability-based harassment that you encountered in social VR and describe what happened? Please feel free to share as much as you're comfortable with.
        \begin{itemize}
            \item In what social context did the harassment happen? 
            \item How many people were involved in the incident? Do you know who they are?
            \item What did the harasser do?
            \item Did that incident impact you afterwards?
                \begin{itemize}
                    \item If so, how?
                \end{itemize}
        \end{itemize}
    \item How did you deal with that harassment incident?
        \begin{itemize}
            \item Did you anticipate it happening and do anything beforehand to try to prevent it?
                \begin{itemize}
                    \item If so, what did you do? And why did you do that?
                \end{itemize}
            \item Did you take any immediate actions?
                \begin{itemize}
                    \item Why or why not?
                \end{itemize}
            \item How did you cope with the experiences afterwards?
        \end{itemize}
    \item Compared to social VR, how do you mitigate disability-based harassment in real life?
\end{enumerate}

\noindent \textbf{Part 2: Understanding Proxemics \& Harassment}
\noindent In this session, we will explore how distance between people in social VR can affect harassment perception. We are going to ask you to put on a VR headset and see three short scenarios, which will show three different types of disability-based harassment at different distances in a social VR space. Please note that these scenes are fictional, and they are designed to help you better understand the different types of harassment and get a spatial sense in social VR. After each scenario, I’ll ask you some follow-up questions. Please know that you don’t have to share anything you’re not comfortable with, and you can pause or exit the study at any time. 

\noindent \textit{Introduce the Four Proximity Layers and Three Types of Harassment Examples.} Now I am going to quickly go through the three types of disability-based harassment that you will see in that environment:
\begin{itemize}
    \item \textbf{Verbal harassment} refers to the use of ableist spoken language to insult, demean, or target someone based on their disability, such as keep calling a wheelchair user ``cripple''; 
    \item \textbf{Physical harassment} refers to the unwanted physical actions in virtual space, such as pushing a PWD’s virtual wheelchair without consent and snatching the walking cane;
    \item \textbf{Environmental harassment} is the display of harmful graphic content in a shared virtual space, such as intentionally using an avatar with stereotypical portraits of disabilities to mimic one’s disabilities.
\end{itemize}

\noindent For each type of harassment, you will see them in four different spaces. This is to help us understand how the experience might feel in each space. We will start with verbal harassment. You will hear the harasser avatar saying `Hey, there is a disabled there. Why there is a disabled?' Before demonstrating, we want to remind you that you can exit the study at any time without penalty. Are you sure you want to continue?  (If yes, ask the following questions:)

\begin{enumerate}
    \item \textbf{Public Space:} How safe or unsafe do you feel when encountering this behavior in the public space? Please rate the sense of safety on a scale 1 to 5, with 1 being very unsafe, and 5 being very safe.
    \begin{itemize}
        \item Could you explain your rating? Why do you feel safe/unsafe?
        \item What makes you feel unsafe?
    \end{itemize}
    \item \textbf{Social Space:} How safe or unsafe do you feel when encountering this behavior in the social space? Please rate the sense of safety on a scale 1 to 5, with 1 being very unsafe, and 5 being very safe.
    \begin{itemize}
        \item Could you explain your rating? Why do you feel safe/unsafe?
    \item What makes you feel unsafe?
    \item Compared to social space, I noticed that your rating has (not) changed, why is that?
    \end{itemize}
    \item \textbf{Personal Space:} How safe or unsafe do you feel when encountering this behavior in the personal space? Please rate the sense of safety on a scale 1 to 5, with 1 being very unsafe, and 5 being very safe.
    \begin{itemize}
        \item Could you explain your rating? Why do you feel safe/unsafe?
    \item What makes you feel unsafe?
    \item Compared to social space, I noticed that your rating has (not) changed, why is that? 
    \end{itemize}
    \item \textbf{Intimate Space:} How safe or unsafe do you feel when encountering this behavior in the intimate space? Please rate the sense of safety on a scale 1 to 5, with 1 being very unsafe, and 5 being very safe.
    \begin{itemize}
        \item Could you explain your rating? Why do you feel safe/unsafe?
        \item What makes you feel unsafe?
        \item Compared to social space, I noticed that your rating has (not) changed, why is that? 
    \end{itemize}
    \item \textbf{Cross-space Reflection:} At which space did you start to feel uncomfortable or threatened by this harassment?
    \begin{itemize}
            \item Could you explain why?
            \item What made it different from other spaces?
        \end{itemize}
\end{enumerate}

\noindent (Repeat the procedure for Physical and Environmental Harassment)

\newpage
\noindent \textbf{Part 3: Co-design Protection Mechanisms}

\noindent \textit{Step 1: General Brainstorming.}
We will start with brainstorming some general ideas. We will ask you questions about what actions you want to take at each to protect yourself from disability-based harassment. Please don't limit by existing technologies or social VR features. Instead, try to imagine the ideal solutions and focus on what you wish social VR can do. We encourage you to be as creative as possible.

\medskip
\noindent \textit{For each type of harassment:}
\begin{enumerate}
    \item In public space, would you like any support to protect yourself from this harassment?
        \begin{itemize}
            \item Why or why not?
            \item What support do you prefer?  
            \item Do you think this space should have ways to protect you from harassment before it happens?
                        \begin{itemize}
                            \item Why or why not?
                            \item If yes, what should it do?
                        \end{itemize}
            \item If something harmful happened in this space, would you want any support to help you respond?
                        \begin{itemize}
                            \item Why or why not?
                            \item If yes, what support would you like?
                        \end{itemize}
                \end{itemize}
    \item How about in the social space? 
        \begin{itemize}
            \item What support do you prefer? Why or why not?
                    \item Do you think this space should have ways to protect you from harassment before it happens?
                        \begin{itemize}
                            \item Why or why not?
                            \item If yes, what should it do?
                        \end{itemize}
                    \item If something harmful happened in this space, would you want any support to help you respond?
                        \begin{itemize}
                            \item Why or why not?
                            \item If yes, what support would you like?
                        \end{itemize}
                    \item How do your preferred actions or support may differ from prior space? Why?
                \end{itemize}
    \item How about in the personal space? 
        \begin{itemize}
            \item What support do you prefer? Why or why not?
                    \item Should social VR do anything before harassment happens?
                        \begin{itemize}
                            \item Why do you think so?
                            \item What should it do?
                        \end{itemize}
                    \item If something harmful already happened in this space, would you want any support to help you respond?
                        \begin{itemize}
                            \item If yes, why?
                            \item What support would you like?
                        \end{itemize}
                    \item How do your preferred actions or support may differ from prior space? Why?
                \end{itemize}
    \item How about in the intimate space?
        \begin{itemize}
            \item What support do you prefer? Why or why not?
                \item Should social VR do anything before harassment happens?
                        \begin{itemize}
                            \item Why do you think so?
                            \item What should it do?
                        \end{itemize}
                    \item If something harmful already happened in this space, would you want any support to help you respond?
                        \begin{itemize}
                            \item If yes, why?
                            \item What support would you like?
                        \end{itemize}
                    \item How do your preferred actions or support may differ from prior space? Why?
                \end{itemize}
\end{enumerate}

\noindent \textit{Step 2: Feedback on Design Probes.} Thank you for sharing your ideas with us. We also come up with four types of protection mechanisms that we would like to discuss with you. We will ask your feedback on how these mechanisms could be effective. Please note that these are just preliminary design examples meant to help us explore how different VR spaces can protect against disability-based harassment. Please feel free to provide any comments on what you like or dislike about them as well as what you think would be a better design.

\subsubsection*{\textbf{Inform Mechanism}} The first mechanism communicates the meaning of your disability signifiers (e.g., virtual wheelchair, virtual cane) to others, preventing potential misunderstandings that lead to inappropriate behaviors. For example, a tool-tip floating on top of your avatar saying, ``My avatar uses a walking cane to represent I am a person with mobility disabilities in real life.'' As you experience it, please think out loud and share any reflections, thoughts, or questions that come to mind.

\begin{enumerate}
    \item Do you think such designs can be effective in preventing verbal harassment in social VR?
        \begin{itemize}
            \item If yes, how can it be effective?
                \begin{itemize}
                    \item Which space would it be most effective? Why?
                \end{itemize}
            \item If no, why not?
        \end{itemize}
    \item What do you like or dislike about such designs?
        \begin{itemize}
            \item Could you explain why?
            \item How would you like to further improve them to better inform others about your disability signifiers on avatars?
        \end{itemize}
    \item Let's talk about the specific design details:
        \begin{itemize}
            \item At which space this mechanism should be triggered, if at all? Why?
            \item Should it work across multiple spaces, or just one? Why?
                \begin{itemize}
                    \item If multiple, what are those spaces? How can it be adaptive across different spaces?
                \end{itemize}
            \item How should this mechanism be triggered in that space(s)? (prompt: should it be automatic or controlled by you? Or be there all the time? Why?)
            \item What kind of information should be conveyed in this mechanism to better inform others and avoid misconception to avatars with disability signifiers?
                \begin{itemize}
                    \item What levels of details feel comfortable to you? (prompt: should it name your specific disability or just the general disabilities? Or just mention the AT you use if there is any? Should it specify why you represent it via avatars or not? Should it be specific about you or a certain group of people with disabilities?)
                    \item Can you explain why?
                \end{itemize}
            \item How would you like to convey that information?
        \end{itemize}
    \item What are potential problems this mechanism might have?
    \item How do you want to further improve it?
\end{enumerate}

\noindent {CHECK other harassment:}
\begin{enumerate}
    \item Would it be effective for environmental harassment?
        \begin{itemize}
            \item Can you explain why?
            \item Would your preferred design for an Inform mechanism be different?
                \begin{itemize}
                    \item Why or why not?
                    \item If it would change, how would it change?
                \end{itemize}
        \end{itemize}
    \item Would it be effective for physical harassment?
        \begin{itemize}
            \item Can you explain why?
            \item Would your preferred design for an Inform mechanism be different?
                \begin{itemize}
                    \item Why or why not?
                    \item If it would change, how would it change?
                \end{itemize}
        \end{itemize}
\end{enumerate}

\subsubsection*{\textbf{Educate Mechanism}} This mechanism aims to help others learn inclusive behaviors when given social opportunities, reducing the ableist expressions and challenging harmful stereotypes that often misrepresent people with disabilities. For example, a signboard appears when someone enters a space: 
Please use inclusive language in this space. E.g., ``Cripple'' is an offensive term and should be avoided. Instead, say ``person with mobility disabilities'' or ``wheelchair users''. 

\begin{enumerate}
    \item Do you think this educational design can be effective in preventing verbal harassment in social VR?
        \begin{itemize}
            \item If yes, how can it be effective?
                \begin{itemize}
                    \item Which space would it be most effective? Why?
                \end{itemize}
            \item If no, why not?
        \end{itemize}
    \item What do you like or dislike about such a design?
        \begin{itemize}
            \item Could you explain why?
            \item How would you like to further improve it?
        \end{itemize}
    \item Let's talk about the design details of this mechanism:
        \begin{itemize}
            \item At which space this mechanism should be triggered?
            \item Should it work across multiple spaces, or just one? Why?
                \begin{itemize}
                    \item If multiple, what are those spaces? How can it be adaptive across different spaces?
                \end{itemize}
            \item When Educate Mechanism is activated, would you like to be notified or aware of it?
                \begin{itemize}
                    \item If yes, how would you like to be notified?
                \end{itemize}
            \item How should this mechanism be triggered in that space(s)?
                \begin{itemize}
                    \item How does your preferred way of triggering this mechanism differ from the Inform Mechanism? Why?
                \end{itemize}
            \item What kind of information should be conveyed in this mechanism to better educate people about inclusive behaviors?
                \begin{itemize}
                    \item What would you like to educate people about? (prompt: what terms should be avoided? What questions are acceptable/not acceptable? Correct bias or stereotypes about people with disabilities?)
                    \item Can you explain your considerations?
                \end{itemize}
            \item How should the reminder be presented?
                \begin{itemize}
                    \item How does your preference differ from Inform Mechanism if there is any? Why?
                \end{itemize}
        \end{itemize}
    \item What are potential problems this mechanism might have?
    \item How do you want to further improve this mechanism?
\end{enumerate}

\noindent {CHECK other harassment:}
\begin{enumerate}
    \item Would it be effective for environmental harassment?
        \begin{itemize}
            \item Why or why not?
            \item Would your preferred design for an Educate Mechanism be different?
                \begin{itemize}
                    \item Why or why not?
                    \item If it would change, how would it change?
                \end{itemize}
        \end{itemize}
    \item Would it be effective for physical harassment?
        \begin{itemize}
            \item Why or why not?
            \item Would your preferred design for this mechanism be different?
                \begin{itemize}
                    \item Why or why not?
                    \item If it would change, how would it change?
                \end{itemize}
        \end{itemize}
\end{enumerate}

\subsubsection*{\textbf{Consent Mechanism}} This mechanism is about consent, which asks for your permission before anyone can touch or interact with you and your disability signifiers, aiming to protect you from unwanted physical contact or unsolicited help. For example, when someone attempts to touch your wheelchair, a virtual shield appears around your avatar and displays: Ask for my consent.

\begin{enumerate}
    \item Do you think this mechanism can be effective in preventing verbal harassment in social VR?
        \begin{itemize}
            \item If yes, how can it be effective?
                \begin{itemize}
                    \item Which space would it be most effective? Why?
                \end{itemize}
            \item If no, why not?
        \end{itemize}
    \item What do you like or dislike about such a design?
        \begin{itemize}
            \item Could you explain why?
            \item How would you like to further improve it for consent purposes?
        \end{itemize}
    \item Let's talk about the design details:
        \begin{itemize}
            \item At which space this mechanism should be triggered?
            \item Should it work across multiple spaces, or just one? Why?
            \item How should this mechanism be triggered in that space(s)?
            \item What information should be included in the Consent mechanism? (prompt: send a request? expectations of how to interact with the AT? Consequences of not following the expectations? Any other ideas?)
                \begin{itemize}
                    \item Can you explain your considerations?
                \end{itemize}
            \item How would you like to consent or be asked for consent? (prompt: by initiating a conversation with you and asking you verbally? Have a button, and they ask for consent by clicking it?)
                \begin{itemize}
                    \item Can you explain your preferences?
                \end{itemize}
        \end{itemize}
    \item What are potential problems this mechanism might have?
    \item How do you want to further improve this mechanism?
\end{enumerate}

\noindent {CHECK other harassment:}
\begin{enumerate}
    \item Would it be effective for environmental harassment?
        \begin{itemize}
            \item Why or why not?
            \item Would your preferred design for this mechanism be different?
                \begin{itemize}
                    \item Why or why not?
                    \item If it would change, how would it change?
                \end{itemize}
        \end{itemize}
    \item Would it be effective for physical harassment?
        \begin{itemize}
            \item Why or why not?
            \item Would your preferred design for this mechanism be different?
                \begin{itemize}
                    \item Why or why not?
                    \item If it would change, how would it change?
                \end{itemize}
        \end{itemize}
\end{enumerate}

\subsubsection*{\textbf{Defense Mechanism}} This mechanism provides immediate support to help you respond to harassment and protect yourself from further harm. For example, if someone keeps attempting to invade your space and touch your avatar, their avatar will be temporarily frozen, and you receive a prompt asking if you want to report them.

\begin{enumerate}
    \item Do you think this mechanism can be effective in preventing verbal harassment in social VR?
        \begin{itemize}
            \item If yes, how can it be effective?
                \begin{itemize}
                    \item Which space would it be most effective? Why?
                \end{itemize}
        \end{itemize}
    \item What do you like or dislike about such a design?
        \begin{itemize}
            \item Could you explain why?
            \item How would you like to further improve it for consent purposes?
        \end{itemize}
    \item Let's talk about the design details:
        \begin{itemize}
            \item Which space(s) should have this mechanism?
            \item Should it work across multiple spaces, or just one?
                \begin{itemize}
                    \item If multiple, how can it be adaptive across different spaces?
                    \item Can you explain your preferences?
                \end{itemize}
            \item How should this mechanism be triggered? (prompt: When detecting any other avatars collide with your avatar? auto popup when anyone is in your space? Controlled by you?)
                \begin{itemize}
                    \item Can you explain your design preference?
                \end{itemize}
            \item What features would you like for Defense Mechanism? Why? (prompt: Mute? Report? remind to not limited by current features)
        \end{itemize}
    \item What are potential problems this mechanism might have?
    \item How do you want to further improve this mechanism?
\end{enumerate}

\noindent {CHECK other harassment:}
\begin{enumerate}
    \item Would it be effective for environmental harassment?
        \begin{itemize}
            \item Why or why not?
            \item Would your preferred design for this mechanism be different?
                \begin{itemize}
                    \item Why or why not?
                    \item If it would change, how would it change?
                \end{itemize}
        \end{itemize}
    \item Would it be effective for physical harassment?
        \begin{itemize}
            \item Why or why not?
            \item Would your preferred design for this mechanism be different?
                \begin{itemize}
                    \item Why or why not?
                    \item If it would change, how would it change?
                \end{itemize}
        \end{itemize}
\end{enumerate}


%% file: main.bbl

\begin{thebibliography}{80}


\ifx \showCODEN    \undefined \def \showCODEN     #1{\unskip}     \fi
\ifx \showDOI      \undefined \def \showDOI       #1{#1}\fi
\ifx \showISBNx    \undefined \def \showISBNx     #1{\unskip}     \fi
\ifx \showISBNxiii \undefined \def \showISBNxiii  #1{\unskip}     \fi
\ifx \showISSN     \undefined \def \showISSN      #1{\unskip}     \fi
\ifx \showLCCN     \undefined \def \showLCCN      #1{\unskip}     \fi
\ifx \shownote     \undefined \def \shownote      #1{#1}          \fi
\ifx \showarticletitle \undefined \def \showarticletitle #1{#1}   \fi
\ifx \showURL      \undefined \def \showURL       {\relax}        \fi
\providecommand\bibfield[2]{#2}
\providecommand\bibinfo[2]{#2}
\providecommand\natexlab[1]{#1}
\providecommand\showeprint[2][]{arXiv:#2}

\bibitem[Abhinaya et~al\mbox{.}(2024)]%
        {abhinaya2024enabling}
\bibfield{author}{\bibinfo{person}{SB Abhinaya}, \bibinfo{person}{Aafaq Sabir}, {and} \bibinfo{person}{Anupam Das}.} \bibinfo{year}{2024}\natexlab{}.
\newblock \showarticletitle{Enabling Developers, Protecting Users: Investigating Harassment and Safety in $\{$VR$\}$}. In \bibinfo{booktitle}{\emph{33rd USENIX Security Symposium (USENIX Security 24)}}. \bibinfo{pages}{6561--6578}.
\newblock


\bibitem[Angerbauer et~al\mbox{.}(2024)]%
        {angerbauer2024_isit?}
\bibfield{author}{\bibinfo{person}{Katrin Angerbauer}, \bibinfo{person}{Phoenix Van~Wagoner}, \bibinfo{person}{Tim Halach}, \bibinfo{person}{Jonas Vogelsang}, \bibinfo{person}{Natalie Hube}, \bibinfo{person}{Andria Smith}, \bibinfo{person}{Ksenia Keplinger}, {and} \bibinfo{person}{Michael Sedlmair}.} \bibinfo{year}{2024}\natexlab{}.
\newblock \showarticletitle{Is it Part of Me? Exploring Experiences of Inclusive Avatar Use For Visible and Invisible Disabilities in Social VR}. In \bibinfo{booktitle}{\emph{Proceedings of the 26th International ACM SIGACCESS Conference on Computers and Accessibility}}. \bibinfo{pages}{1--15}.
\newblock


\bibitem[{Anti-Defamation League}(2025)]%
        {ADL_OnlineHateHarassmentSafetyGuide}
\bibfield{author}{\bibinfo{person}{{Anti-Defamation League}}.} \bibinfo{year}{2025}\natexlab{}.
\newblock \bibinfo{title}{Online Hate \& Harassment Safety Guide}.
\newblock \bibinfo{howpublished}{\url{https://www.adl.org/online-hate-and-harassment-safety-guide}}.
\newblock
\newblock
\shownote{Accessed: 2025-09-08}.


\bibitem[Bailenson et~al\mbox{.}(2003)]%
        {bailenson2003interpersonal}
\bibfield{author}{\bibinfo{person}{Jeremy~N Bailenson}, \bibinfo{person}{Jim Blascovich}, \bibinfo{person}{Andrew~C Beall}, {and} \bibinfo{person}{Jack~M Loomis}.} \bibinfo{year}{2003}\natexlab{}.
\newblock \showarticletitle{Interpersonal distance in immersive virtual environments}.
\newblock \bibinfo{journal}{\emph{Personality and social psychology bulletin}} \bibinfo{volume}{29}, \bibinfo{number}{7} (\bibinfo{year}{2003}), \bibinfo{pages}{819--833}.
\newblock


\bibitem[Barnes(1992)]%
        {barnes1992disabling}
\bibfield{author}{\bibinfo{person}{Colin Barnes}.} \bibinfo{year}{1992}\natexlab{}.
\newblock \showarticletitle{Disabling imagery and the media}.
\newblock \bibinfo{journal}{\emph{An Exploration of the Principles for Media Representations of Disabled People. The First in a Series of Reports. Halifax}}  \bibinfo{volume}{28} (\bibinfo{year}{1992}).
\newblock


\bibitem[Blackwell et~al\mbox{.}(2019)]%
        {blackwell2019harassment}
\bibfield{author}{\bibinfo{person}{Lindsay Blackwell}, \bibinfo{person}{Nicole Ellison}, \bibinfo{person}{Natasha Elliott-Deflo}, {and} \bibinfo{person}{Raz Schwartz}.} \bibinfo{year}{2019}\natexlab{}.
\newblock \showarticletitle{Harassment in social virtual reality: Challenges for platform governance}.
\newblock \bibinfo{journal}{\emph{Proceedings of the ACM on Human-Computer Interaction}} \bibinfo{volume}{3}, \bibinfo{number}{CSCW} (\bibinfo{year}{2019}), \bibinfo{pages}{1--25}.
\newblock


\bibitem[Braun and Clarke(2006)]%
        {braun2006theme}
\bibfield{author}{\bibinfo{person}{Virginia Braun} {and} \bibinfo{person}{Victoria Clarke}.} \bibinfo{year}{2006}\natexlab{}.
\newblock \showarticletitle{Using thematic analysis in psychology}.
\newblock \bibinfo{journal}{\emph{Qualitative research in psychology}} \bibinfo{volume}{3}, \bibinfo{number}{2} (\bibinfo{year}{2006}), \bibinfo{pages}{77--101}.
\newblock


\bibitem[Chen et~al\mbox{.}(2025)]%
        {Chen2025DemocraticModeration}
\bibfield{author}{\bibinfo{person}{Qijia Chen}, \bibinfo{person}{Qunfang Wu}, {and} \bibinfo{person}{Giulio Jacucci}.} \bibinfo{year}{2025}\natexlab{}.
\newblock \showarticletitle{Democratic Moderation: Exploring the Use and Perception of Votekicking in Social Virtual Reality}. In \bibinfo{booktitle}{\emph{Proceedings of the 2025 CHI Conference on Human Factors in Computing Systems}} (Yokohama, Japan) \emph{(\bibinfo{series}{CHI '25})}. \bibinfo{publisher}{Association for Computing Machinery}, \bibinfo{address}{New York, NY, USA}, Article \bibinfo{articleno}{493}, \bibinfo{numpages}{18}~pages.
\newblock
\showISBNx{979-8-4007-1394-1}
\urldef\tempurl%
\url{https://doi.org/10.1145/3706598.3713577}
\showDOI{\tempurl}


\bibitem[Cohen(2013)]%
        {cohen2013statistical}
\bibfield{author}{\bibinfo{person}{Jacob Cohen}.} \bibinfo{year}{2013}\natexlab{}.
\newblock \bibinfo{booktitle}{\emph{Statistical power analysis for the behavioral sciences}}.
\newblock \bibinfo{publisher}{routledge}.
\newblock


\bibitem[Corrigan and Watson(2002)]%
        {corrigan2002paradox}
\bibfield{author}{\bibinfo{person}{Patrick~W. Corrigan} {and} \bibinfo{person}{Amy~C. Watson}.} \bibinfo{year}{2002}\natexlab{}.
\newblock \showarticletitle{The paradox of self-stigma and mental illness}.
\newblock \bibinfo{journal}{\emph{Clinical Psychology: Science and Practice}} \bibinfo{volume}{9}, \bibinfo{number}{1} (\bibinfo{year}{2002}), \bibinfo{pages}{35--53}.
\newblock
\urldef\tempurl%
\url{https://doi.org/10.1093/clipsy.9.1.35}
\showDOI{\tempurl}


\bibitem[Creswell and Miller(2000)]%
        {creswell_determining_2000}
\bibfield{author}{\bibinfo{person}{John~W. Creswell} {and} \bibinfo{person}{Dana~L. Miller}.} \bibinfo{year}{2000}\natexlab{}.
\newblock \showarticletitle{Determining validity in qualitative inquiry}.
\newblock \bibinfo{journal}{\emph{Theory Into Practice}} \bibinfo{volume}{39}, \bibinfo{number}{3} (\bibinfo{year}{2000}), \bibinfo{pages}{124--130}.
\newblock
\showISSN{0040-5841}
\urldef\tempurl%
\url{https://doi.org/10.1207/s15430421tip3903_2}
\showDOI{\tempurl}


\bibitem[Czopp et~al\mbox{.}(2006)]%
        {czopp2006standing}
\bibfield{author}{\bibinfo{person}{Alexander~M. Czopp}, \bibinfo{person}{Margo~J. Monteith}, {and} \bibinfo{person}{Aimee~Y. Mark}.} \bibinfo{year}{2006}\natexlab{}.
\newblock \showarticletitle{Standing up for a change: Reducing bias through interpersonal confrontation}.
\newblock \bibinfo{journal}{\emph{Journal of Personality and Social Psychology}} \bibinfo{volume}{90}, \bibinfo{number}{5} (\bibinfo{year}{2006}), \bibinfo{pages}{784--803}.
\newblock
\urldef\tempurl%
\url{https://doi.org/10.1037/0022-3514.90.5.784}
\showDOI{\tempurl}


\bibitem[Daley and Rappolt-Schlichtmann(2018)]%
        {daley2018stigma}
\bibfield{author}{\bibinfo{person}{Samantha~G Daley} {and} \bibinfo{person}{Gabrielle Rappolt-Schlichtmann}.} \bibinfo{year}{2018}\natexlab{}.
\newblock \showarticletitle{Stigma consciousness among adolescents with learning disabilities: Considering individual experiences of being stereotyped}.
\newblock \bibinfo{journal}{\emph{Learning Disability Quarterly}} \bibinfo{volume}{41}, \bibinfo{number}{4} (\bibinfo{year}{2018}), \bibinfo{pages}{200--212}.
\newblock


\bibitem[de~Rosa and Mannarini(2021)]%
        {deRosa2021Covid}
\bibfield{author}{\bibinfo{person}{A.~S. de Rosa} {and} \bibinfo{person}{T. Mannarini}.} \bibinfo{year}{2021}\natexlab{}.
\newblock \showarticletitle{{COVID-19} as an ``Invisible Other'' and Socio-Spatial Distancing within a One-Metre Individual Bubble}.
\newblock \bibinfo{journal}{\emph{Urban Design International}} \bibinfo{volume}{26}, \bibinfo{number}{4} (\bibinfo{year}{2021}), \bibinfo{pages}{370--390}.
\newblock
\urldef\tempurl%
\url{https://doi.org/10.1057/s41289-021-00151-z}
\showDOI{\tempurl}


\bibitem[Eisenberg et~al\mbox{.}(2015)]%
        {eisenberg2015vulnerable}
\bibfield{author}{\bibinfo{person}{Marla~E Eisenberg}, \bibinfo{person}{Amy~L Gower}, \bibinfo{person}{Barbara~J McMorris}, {and} \bibinfo{person}{Michaela~M Bucchianeri}.} \bibinfo{year}{2015}\natexlab{}.
\newblock \showarticletitle{Vulnerable bullies: Perpetration of peer harassment among youths across sexual orientation, weight, and disability status}.
\newblock \bibinfo{journal}{\emph{American journal of public health}} \bibinfo{volume}{105}, \bibinfo{number}{9} (\bibinfo{year}{2015}), \bibinfo{pages}{1784--1791}.
\newblock


\bibitem[Emerson et~al\mbox{.}(2021)]%
        {emerson2021loneliness}
\bibfield{author}{\bibinfo{person}{Eric Emerson}, \bibinfo{person}{Nicola Fortune}, \bibinfo{person}{Gwynnyth Llewellyn}, {and} \bibinfo{person}{Roger Stancliffe}.} \bibinfo{year}{2021}\natexlab{}.
\newblock \showarticletitle{Loneliness, social support, social isolation and wellbeing among working age adults with and without disability: Cross-sectional study}.
\newblock \bibinfo{journal}{\emph{Disability and health journal}} \bibinfo{volume}{14}, \bibinfo{number}{1} (\bibinfo{year}{2021}), \bibinfo{pages}{100965}.
\newblock


\bibitem[Fiani et~al\mbox{.}(2024)]%
        {fiani2024pikachu}
\bibfield{author}{\bibinfo{person}{Cristina Fiani}, \bibinfo{person}{Robin Bretin}, \bibinfo{person}{Shaun~Alexander Macdonald}, \bibinfo{person}{Mohamed Khamis}, {and} \bibinfo{person}{Mark Mcgill}.} \bibinfo{year}{2024}\natexlab{}.
\newblock \showarticletitle{``Pikachu would electrocute people who are misbehaving': Expert, Guardian and Child Perspectives on Automated Embodied Moderators for Safeguarding Children in Social Virtual Reality}. In \bibinfo{booktitle}{\emph{Proceedings of the 2024 CHI Conference on Human Factors in Computing Systems}}. \bibinfo{pages}{1--23}.
\newblock


\bibitem[Fiani et~al\mbox{.}(2023)]%
        {fiani_bigbuddy_2023}
\bibfield{author}{\bibinfo{person}{Cristina Fiani}, \bibinfo{person}{Robin Bretin}, \bibinfo{person}{Mark Mcgill}, {and} \bibinfo{person}{Mohamed Khamis}.} \bibinfo{year}{2023}\natexlab{}.
\newblock \showarticletitle{Big Buddy: Exploring Child Reactions and Parental Perceptions towards a Simulated Embodied Moderating System for Social Virtual Reality}. In \bibinfo{booktitle}{\emph{Proceedings of the 22nd Annual ACM Interaction Design and Children Conference}} (Chicago, IL, USA) \emph{(\bibinfo{series}{IDC '23})}. \bibinfo{publisher}{Association for Computing Machinery}, \bibinfo{address}{New York, NY, USA}, \bibinfo{pages}{1–13}.
\newblock
\showISBNx{9798400701313}
\urldef\tempurl%
\url{https://doi.org/10.1145/3585088.3589374}
\showDOI{\tempurl}


\bibitem[Freeman et~al\mbox{.}(2025)]%
        {freeman_comforting_2025}
\bibfield{author}{\bibinfo{person}{Guo Freeman}, \bibinfo{person}{Kelsea Schulenberg}, \bibinfo{person}{Lingyuan Li}, \bibinfo{person}{Ruchi Panchanadikar}, {and} \bibinfo{person}{Nathan McNeese}.} \bibinfo{year}{2025}\natexlab{}.
\newblock \showarticletitle{``Comforting and Small Like a House Cat, Big and Intimidating Like a Bodyguard'': How Women Perceive and Envision {AI} Companions as a New Harassment Mitigation Approach in Social Virtual Reality}. In \bibinfo{booktitle}{\emph{Proceedings of the CHI Conference on Human Factors in Computing Systems}} (Yokohama, Japan) \emph{(\bibinfo{series}{CHI '25})}. \bibinfo{publisher}{Association for Computing Machinery}, \bibinfo{address}{New York, NY, USA}, \bibinfo{pages}{1--16}.
\newblock
\showISBNx{9798400713064}
\urldef\tempurl%
\url{https://doi.org/10.1145/3706598.3713473}
\showDOI{\tempurl}


\bibitem[Freeman et~al\mbox{.}(2022)]%
        {freeman2022disturbing}
\bibfield{author}{\bibinfo{person}{Guo Freeman}, \bibinfo{person}{Samaneh Zamanifard}, \bibinfo{person}{Divine Maloney}, {and} \bibinfo{person}{Dane Acena}.} \bibinfo{year}{2022}\natexlab{}.
\newblock \showarticletitle{Disturbing the peace: Experiencing and mitigating emerging harassment in social virtual reality}.
\newblock \bibinfo{journal}{\emph{Proceedings of the ACM on Human-Computer Interaction}} \bibinfo{volume}{6}, \bibinfo{number}{CSCW1} (\bibinfo{year}{2022}), \bibinfo{pages}{1--30}.
\newblock


\bibitem[Freeman et~al\mbox{.}(2020)]%
        {freeman2020my}
\bibfield{author}{\bibinfo{person}{Guo Freeman}, \bibinfo{person}{Samaneh Zamanifard}, \bibinfo{person}{Divine Maloney}, {and} \bibinfo{person}{Alexandra Adkins}.} \bibinfo{year}{2020}\natexlab{}.
\newblock \showarticletitle{My body, my avatar: How people perceive their avatars in social virtual reality}. In \bibinfo{booktitle}{\emph{Extended Abstracts of the 2020 CHI Conference on Human Factors in Computing Systems}}. \bibinfo{pages}{1--8}.
\newblock


\bibitem[Goffman(1963)]%
        {goffman1963stigma}
\bibfield{author}{\bibinfo{person}{Erving Goffman}.} \bibinfo{year}{1963}\natexlab{}.
\newblock \bibinfo{booktitle}{\emph{Stigma: Notes on the Management of Spoiled Identity}}.
\newblock \bibinfo{publisher}{Prentice-Hall}.
\newblock


\bibitem[Grau et~al\mbox{.}(1985)]%
        {restrainingorder_1985}
\bibfield{author}{\bibinfo{person}{Janice Grau}, \bibinfo{person}{Jeffrey Fagan}, {and} \bibinfo{person}{Sandra Wexler}.} \bibinfo{year}{1985}\natexlab{}.
\newblock \showarticletitle{Restraining Orders for Battered Women: Issues of Access and Efficacy}.
\newblock In \bibinfo{booktitle}{\emph{Criminal Justice Politics and Women: The Aftermath of Legally Mandated Change}}, \bibfield{editor}{\bibinfo{person}{Claudine SchWeber} {and} \bibinfo{person}{Clarice Feinman}} (Eds.). \bibinfo{publisher}{Haworth Press}, \bibinfo{address}{New York, NY}, \bibinfo{pages}{13--28}.
\newblock
\showISBNx{0866563644}
\urldef\tempurl%
\url{https://doi.org/10.4324/9781315860114-3}
\showDOI{\tempurl}


\bibitem[Gray et~al\mbox{.}(2024)]%
        {gray2024_book}
\bibfield{author}{\bibinfo{person}{Joanne~E. Gray}, \bibinfo{person}{Marcus Carter}, {and} \bibinfo{person}{Ben Egliston}.} \bibinfo{year}{2024}\natexlab{}.
\newblock \bibinfo{booktitle}{\emph{Content Harms in Social VR: Abuse, Misinformation, Platform Cultures and Moderation}}.
\newblock \bibinfo{publisher}{Springer Nature Switzerland}, \bibinfo{address}{Cham}, \bibinfo{pages}{11--22}.
\newblock
\showISBNx{978-3-031-61831-4}
\urldef\tempurl%
\url{https://doi.org/10.1007/978-3-031-61831-4_2}
\showDOI{\tempurl}


\bibitem[Greenberg et~al\mbox{.}(2011)]%
        {Greenberg2011ProxemicInteractions}
\bibfield{author}{\bibinfo{person}{Saul Greenberg}, \bibinfo{person}{Nicolai Marquardt}, \bibinfo{person}{Till Ballendat}, \bibinfo{person}{Rob Diaz{-}Marino}, {and} \bibinfo{person}{Miaosen Wang}.} \bibinfo{year}{2011}\natexlab{}.
\newblock \showarticletitle{Proxemic Interactions: The New Ubicomp?}
\newblock \bibinfo{journal}{\emph{interactions}} \bibinfo{volume}{18}, \bibinfo{number}{1} (\bibinfo{date}{Jan.} \bibinfo{year}{2011}), \bibinfo{pages}{42--50}.
\newblock
\urldef\tempurl%
\url{https://doi.org/10.1145/1897239.1897250}
\showDOI{\tempurl}


\bibitem[Gualano et~al\mbox{.}(2024)]%
        {Gualano_invisible_full2024}
\bibfield{author}{\bibinfo{person}{Ria~J. Gualano}, \bibinfo{person}{Lucy Jiang}, \bibinfo{person}{Kexin Zhang}, \bibinfo{person}{Tanisha Shende}, \bibinfo{person}{Andrea~Stevenson Won}, {and} \bibinfo{person}{Shiri Azenkot}.} \bibinfo{year}{2024}\natexlab{}.
\newblock \showarticletitle{“I Try to Represent Myself as I Am”: Self-Presentation Preferences of People with Invisible Disabilities through Embodied Social VR Avatars}. In \bibinfo{booktitle}{\emph{Proceedings of the 26th International ACM SIGACCESS Conference on Computers and Accessibility}} (St. John's, NL, Canada) \emph{(\bibinfo{series}{ASSETS '24})}. \bibinfo{publisher}{Association for Computing Machinery}, \bibinfo{address}{New York, NY, USA}, Article \bibinfo{articleno}{72}, \bibinfo{numpages}{15}~pages.
\newblock
\showISBNx{9798400706776}
\urldef\tempurl%
\url{https://doi.org/10.1145/3663548.3675620}
\showDOI{\tempurl}


\bibitem[Hall(1966)]%
        {hall_1966_hidden}
\bibfield{author}{\bibinfo{person}{Edward~T. Hall}.} \bibinfo{year}{1966}\natexlab{}.
\newblock \bibinfo{booktitle}{\emph{The Hidden Dimension}}.
\newblock \bibinfo{publisher}{Doubleday}, \bibinfo{address}{New York}.
\newblock


\bibitem[Hall(2009)]%
        {hall2009social}
\bibfield{author}{\bibinfo{person}{Sarah~A Hall}.} \bibinfo{year}{2009}\natexlab{}.
\newblock \showarticletitle{The social inclusion of people with disabilities: a qualitative meta-analysis.}
\newblock \bibinfo{journal}{\emph{Journal of ethnographic \& qualitative research}} \bibinfo{volume}{3}, \bibinfo{number}{3} (\bibinfo{year}{2009}).
\newblock


\bibitem[Heung et~al\mbox{.}(2024)]%
        {heung2024vulnerable}
\bibfield{author}{\bibinfo{person}{Sharon Heung}, \bibinfo{person}{Lucy Jiang}, \bibinfo{person}{Shiri Azenkot}, {and} \bibinfo{person}{Aditya Vashistha}.} \bibinfo{year}{2024}\natexlab{}.
\newblock \showarticletitle{``Vulnerable, Victimized, and Objectified'': Understanding Ableist Hate and Harassment Experienced by Disabled Content Creators on Social Media}. In \bibinfo{booktitle}{\emph{Proceedings of the 2024 CHI Conference on Human Factors in Computing Systems}} \emph{(\bibinfo{series}{CHI '24})}. \bibinfo{publisher}{Association for Computing Machinery}, \bibinfo{address}{New York, NY, USA}, Article \bibinfo{articleno}{744}, \bibinfo{numpages}{19}~pages.
\newblock
\urldef\tempurl%
\url{https://doi.org/10.1145/3613904.3641949}
\showDOI{\tempurl}


\bibitem[Heung et~al\mbox{.}(2025)]%
        {heung2025ignorance}
\bibfield{author}{\bibinfo{person}{Sharon Heung}, \bibinfo{person}{Lucy Jiang}, \bibinfo{person}{Shiri Azenkot}, {and} \bibinfo{person}{Aditya Vashistha}.} \bibinfo{year}{2025}\natexlab{}.
\newblock \showarticletitle{``Ignorance is not Bliss'': Designing Personalized Moderation to Address Ableist Hate on Social Media}. In \bibinfo{booktitle}{\emph{Proceedings of the 2025 CHI Conference on Human Factors in Computing Systems}} \emph{(\bibinfo{series}{CHI '25})}. \bibinfo{publisher}{Association for Computing Machinery}, \bibinfo{address}{New York, NY, USA}, \bibinfo{numpages}{18}~pages.
\newblock
\urldef\tempurl%
\url{https://doi.org/10.1145/3706598.3713997}
\showDOI{\tempurl}


\bibitem[Heung et~al\mbox{.}(2022)]%
        {heung2022nothing}
\bibfield{author}{\bibinfo{person}{Sharon Heung}, \bibinfo{person}{Mahika Phutane}, \bibinfo{person}{Shiri Azenkot}, \bibinfo{person}{Megh Marathe}, {and} \bibinfo{person}{Aditya Vashistha}.} \bibinfo{year}{2022}\natexlab{}.
\newblock \showarticletitle{Nothing Micro About It: Examining Ableist Microaggressions on Social Media}. In \bibinfo{booktitle}{\emph{Proceedings of the 24th International ACM SIGACCESS Conference on Computers and Accessibility}} \emph{(\bibinfo{series}{ASSETS '22})}. \bibinfo{publisher}{Association for Computing Machinery}, \bibinfo{address}{New York, NY, USA}, Article \bibinfo{articleno}{27}, \bibinfo{numpages}{14}~pages.
\newblock
\urldef\tempurl%
\url{https://doi.org/10.1145/3517428.3544801}
\showDOI{\tempurl}


\bibitem[Howard et~al\mbox{.}(2022)]%
        {howard2022ATuse}
\bibfield{author}{\bibinfo{person}{Jonathan Howard}, \bibinfo{person}{Zoe Fisher}, \bibinfo{person}{Andrew~H Kemp}, \bibinfo{person}{Stephen Lindsay}, \bibinfo{person}{Lorna~H Tasker}, {and} \bibinfo{person}{Jeremy~J Tree}.} \bibinfo{year}{2022}\natexlab{}.
\newblock \showarticletitle{Exploring the barriers to using assistive technology for individuals with chronic conditions: a meta-synthesis review}.
\newblock \bibinfo{journal}{\emph{Disability and rehabilitation: Assistive technology}} \bibinfo{volume}{17}, \bibinfo{number}{4} (\bibinfo{year}{2022}), \bibinfo{pages}{390--408}.
\newblock


\bibitem[Hutchinson et~al\mbox{.}(2003)]%
        {hutchinson2003probes}
\bibfield{author}{\bibinfo{person}{Hilary Hutchinson}, \bibinfo{person}{Wendy Mackay}, \bibinfo{person}{Bo Westerlund}, \bibinfo{person}{Benjamin~B Bederson}, \bibinfo{person}{Allison Druin}, \bibinfo{person}{Catherine Plaisant}, \bibinfo{person}{Michel Beaudouin-Lafon}, \bibinfo{person}{St{\'e}phane Conversy}, \bibinfo{person}{Helen Evans}, \bibinfo{person}{Heiko Hansen}, {et~al\mbox{.}}} \bibinfo{year}{2003}\natexlab{}.
\newblock \showarticletitle{Technology probes: inspiring design for and with families}. In \bibinfo{booktitle}{\emph{Proceedings of the SIGCHI conference on Human factors in computing systems}}. \bibinfo{pages}{17--24}.
\newblock


\bibitem[Ji et~al\mbox{.}(2022)]%
        {ji2022vrbubble}
\bibfield{author}{\bibinfo{person}{Tiger~F Ji}, \bibinfo{person}{Brianna Cochran}, {and} \bibinfo{person}{Yuhang Zhao}.} \bibinfo{year}{2022}\natexlab{}.
\newblock \showarticletitle{Vrbubble: Enhancing peripheral awareness of avatars for people with visual impairments in social virtual reality}. In \bibinfo{booktitle}{\emph{Proceedings of the 24th International ACM SIGACCESS Conference on Computers and Accessibility}}. \bibinfo{pages}{1--17}.
\newblock


\bibitem[Joachim and Acorn(2000)]%
        {joachim2000uncertainty}
\bibfield{author}{\bibinfo{person}{Gloria Joachim} {and} \bibinfo{person}{Sonia Acorn}.} \bibinfo{year}{2000}\natexlab{}.
\newblock \showarticletitle{Uncertainty and the stigma of chronic illness}.
\newblock \bibinfo{journal}{\emph{Journal of Advanced Nursing}} \bibinfo{volume}{32}, \bibinfo{number}{3} (\bibinfo{year}{2000}), \bibinfo{pages}{609--616}.
\newblock


\bibitem[Kiesler et~al\mbox{.}(2012)]%
        {kiesler2012regulating}
\bibfield{author}{\bibinfo{person}{Sara Kiesler}, \bibinfo{person}{Robert Kraut}, \bibinfo{person}{Paul Resnick}, {and} \bibinfo{person}{Aniket Kittur}.} \bibinfo{year}{2012}\natexlab{}.
\newblock \showarticletitle{Regulating behavior in online communities}.
\newblock \bibinfo{journal}{\emph{Building successful online communities: Evidence-based social design}}  \bibinfo{volume}{1} (\bibinfo{year}{2012}), \bibinfo{pages}{4--2}.
\newblock


\bibitem[Lazerson(2022)]%
        {lazerson_secure_2022}
\bibfield{author}{\bibinfo{person}{R. Lazerson}.} \bibinfo{year}{2022}\natexlab{}.
\newblock \bibinfo{booktitle}{\emph{A Secure and Equitable Metaverse: Designing Effective Community Guidelines for Social VR}}.
\newblock \bibinfo{type}{{T}echnical {R}eport}. \bibinfo{institution}{Centre for Long-Term Cybersecurity, UC Berkeley}.
\newblock
\urldef\tempurl%
\url{https://cltc.berkeley.edu/wp-content/uploads/2022/11/Secure_Equitable_Metaverse.pdf}
\showURL{%
\tempurl}
\newblock
\shownote{Tech Report}.


\bibitem[Lee et~al\mbox{.}(2026)]%
        {lee2025harassguard}
\bibfield{author}{\bibinfo{person}{Junhee Lee}, \bibinfo{person}{Minseok Kim}, \bibinfo{person}{Hwanjo Heo}, {and} \bibinfo{person}{Jinwoo Kim}.} \bibinfo{year}{2026}\natexlab{}.
\newblock \showarticletitle{HarassGuard: Detecting Harassment Behaviors in Social Virtual Reality with Vision-Language Models}.
\newblock \bibinfo{journal}{\emph{IEEE Transactions on Visualization and Computer Graphics}} (\bibinfo{year}{2026}).
\newblock
\newblock
\shownote{Special Issue on IEEE VR 2026; arXiv:2604.00592}.


\bibitem[Liao et~al\mbox{.}(2025)]%
        {liao2025_proactive_safety}
\bibfield{author}{\bibinfo{person}{Zhehui Liao}, \bibinfo{person}{Hanwen Zhao}, \bibinfo{person}{Ayush Kulkarni}, \bibinfo{person}{Shaan~Singh Chattrath}, {and} \bibinfo{person}{Amy~X. Zhang}.} \bibinfo{year}{2025}\natexlab{}.
\newblock \showarticletitle{Building Proactive and Instant‑Reactive Safety Designs to Address Harassment in Social Virtual Reality}.
\newblock \bibinfo{journal}{\emph{arXiv}} (\bibinfo{year}{2025}).
\newblock
\showeprint[arxiv]{2504.05781}~[cs.HC]
\urldef\tempurl%
\url{https://arxiv.org/abs/2504.05781}
\showURL{%
\tempurl}
\newblock
\shownote{37 pages, 11 figures}.


\bibitem[Lingsom(2008)]%
        {lingsom2008invisibleImpairments}
\bibfield{author}{\bibinfo{person}{Susan Lingsom}.} \bibinfo{year}{2008}\natexlab{}.
\newblock \showarticletitle{Invisible Impairments: Dilemmas of Concealment and Disclosure}.
\newblock \bibinfo{journal}{\emph{Scandinavian Journal of Disability Research}} \bibinfo{volume}{10}, \bibinfo{number}{1} (\bibinfo{year}{2008}), \bibinfo{pages}{2--16}.
\newblock
\urldef\tempurl%
\url{https://doi.org/10.1080/15017410701391567}
\showDOI{\tempurl}


\bibitem[Low(2019)]%
        {Low2019Spikes}
\bibfield{author}{\bibinfo{person}{Harry Low}.} \bibinfo{year}{2019}\natexlab{}.
\newblock \bibinfo{booktitle}{\emph{Spikes — and other ways disabled people combat unwanted touching}}.
\newblock BBC News.
\newblock
\urldef\tempurl%
\url{https://www.bbc.com/news/disability-49584591}
\showURL{%
\tempurl}
\newblock
\shownote{BBC Ouch}.


\bibitem[Mack et~al\mbox{.}(2023)]%
        {mack2023towards}
\bibfield{author}{\bibinfo{person}{Kelly Mack}, \bibinfo{person}{Rai Ching~Ling Hsu}, \bibinfo{person}{Andr{\'e}s Monroy-Hern{\'a}ndez}, \bibinfo{person}{Brian~A Smith}, {and} \bibinfo{person}{Fannie Liu}.} \bibinfo{year}{2023}\natexlab{}.
\newblock \showarticletitle{Towards inclusive avatars: Disability representation in avatar platforms}. In \bibinfo{booktitle}{\emph{Proceedings of the 2023 CHI Conference on Human Factors in Computing Systems}}. \bibinfo{pages}{1--13}.
\newblock


\bibitem[Mack et~al\mbox{.}(2024)]%
        {mack2024t2i}
\bibfield{author}{\bibinfo{person}{Kelly~Avery Mack}, \bibinfo{person}{Rida Qadri}, \bibinfo{person}{Remi Denton}, \bibinfo{person}{Shaun~K Kane}, {and} \bibinfo{person}{Cynthia~L Bennett}.} \bibinfo{year}{2024}\natexlab{}.
\newblock \showarticletitle{“They only care to show us the wheelchair”: disability representation in text-to-image AI models}. In \bibinfo{booktitle}{\emph{Proceedings of the 2024 CHI Conference on Human Factors in Computing Systems}}. \bibinfo{pages}{1--23}.
\newblock


\bibitem[Maloney et~al\mbox{.}(2020)]%
        {maloney2020talking}
\bibfield{author}{\bibinfo{person}{Divine Maloney}, \bibinfo{person}{Guo Freeman}, {and} \bibinfo{person}{Donghee~Yvette Wohn}.} \bibinfo{year}{2020}\natexlab{}.
\newblock \showarticletitle{" Talking without a voice" understanding non-verbal communication in social virtual reality}.
\newblock \bibinfo{journal}{\emph{Proceedings of the ACM on human-computer interaction}} \bibinfo{volume}{4}, \bibinfo{number}{CSCW2} (\bibinfo{year}{2020}), \bibinfo{pages}{1--25}.
\newblock


\bibitem[Manno et~al\mbox{.}(2024)]%
        {Manno2024DisabilityDisclosure}
\bibfield{author}{\bibinfo{person}{Cherie~M. Manno}, \bibinfo{person}{Rachel Glade}, \bibinfo{person}{Lynn~C. Koch}, \bibinfo{person}{Laura~S. Simon}, \bibinfo{person}{Phillip~D. Rumrill}, {and} \bibinfo{person}{Christopher~C. Rosen}.} \bibinfo{year}{2024}\natexlab{}.
\newblock \showarticletitle{Disability disclosure as an impression management technique used in the workplace: A grounded theory investigation}.
\newblock \bibinfo{journal}{\emph{Work}} \bibinfo{volume}{78}, \bibinfo{number}{2} (\bibinfo{year}{2024}), \bibinfo{pages}{219--233}.
\newblock
\urldef\tempurl%
\url{https://doi.org/10.3233/WOR-246007}
\showDOI{\tempurl}


\bibitem[{Meta}(2022)]%
        {Meta2022PersonalBoundary}
\bibfield{author}{\bibinfo{person}{{Meta}}.} \bibinfo{year}{2022}\natexlab{}.
\newblock \bibinfo{title}{Introducing a Personal Boundary for Horizon Worlds and Venues}.
\newblock \bibinfo{howpublished}{\url{https://www.meta.com/blog/introducing-a-personal-boundary-for-horizon-worlds-and-venues/}}.
\newblock
\newblock
\shownote{Meta Blog; accessed 2025-08-12}.


\bibitem[Miron et~al\mbox{.}(2023)]%
        {miron2023onlinedating}
\bibfield{author}{\bibinfo{person}{M. Miron}, \bibinfo{person}{K. Goulet}, \bibinfo{person}{L.~P. Auger}, {et~al\mbox{.}}} \bibinfo{year}{2023}\natexlab{}.
\newblock \showarticletitle{Online Dating for People with Disabilities: A Scoping Review}.
\newblock \bibinfo{journal}{\emph{Sexuality and Disability}}  \bibinfo{volume}{41} (\bibinfo{year}{2023}), \bibinfo{pages}{31--61}.
\newblock
\urldef\tempurl%
\url{https://doi.org/10.1007/s11195-022-09771-x}
\showDOI{\tempurl}


\bibitem[Morris et~al\mbox{.}(2023)]%
        {morris2023don}
\bibfield{author}{\bibinfo{person}{Margaret~E Morris}, \bibinfo{person}{Daniela~K Rosner}, \bibinfo{person}{Paula~S Nurius}, {and} \bibinfo{person}{Hadar~M Dolev}.} \bibinfo{year}{2023}\natexlab{}.
\newblock \showarticletitle{“I don't want to hide behind an avatar”: Self-representation in social VR among women in midlife}. In \bibinfo{booktitle}{\emph{Proceedings of the 2023 ACM designing interactive systems conference}}. \bibinfo{pages}{537--546}.
\newblock


\bibitem[Moustafa and Steed(2018)]%
        {moustafa2018longitudinal}
\bibfield{author}{\bibinfo{person}{Fares Moustafa} {and} \bibinfo{person}{Anthony Steed}.} \bibinfo{year}{2018}\natexlab{}.
\newblock \showarticletitle{A longitudinal study of small group interaction in social virtual reality}. In \bibinfo{booktitle}{\emph{Proceedings of the 24th ACM symposium on virtual reality software and technology}}. \bibinfo{pages}{1--10}.
\newblock


\bibitem[Nario-Redmond et~al\mbox{.}(2019)]%
        {nario2019hostile}
\bibfield{author}{\bibinfo{person}{Michelle~R Nario-Redmond}, \bibinfo{person}{Angela~A Kemerling}, {and} \bibinfo{person}{Arielle Silverman}.} \bibinfo{year}{2019}\natexlab{}.
\newblock \showarticletitle{Hostile, benevolent, and ambivalent ableism: Contemporary manifestations}.
\newblock \bibinfo{journal}{\emph{Journal of Social Issues}} \bibinfo{volume}{75}, \bibinfo{number}{3} (\bibinfo{year}{2019}), \bibinfo{pages}{726--756}.
\newblock


\bibitem[Nario-Redmond et~al\mbox{.}(2013)]%
        {narioredmond2013redefining}
\bibfield{author}{\bibinfo{person}{Michelle~R. Nario-Redmond}, \bibinfo{person}{Jeffrey~G. Noel}, {and} \bibinfo{person}{Emily Fern}.} \bibinfo{year}{2013}\natexlab{}.
\newblock \showarticletitle{Redefining disability, re-imagining the self: Disability identification predicts self-esteem and strategic responses to stigma}.
\newblock \bibinfo{journal}{\emph{Self and Identity}} \bibinfo{volume}{12}, \bibinfo{number}{5} (\bibinfo{year}{2013}), \bibinfo{pages}{468--488}.
\newblock
\urldef\tempurl%
\url{https://doi.org/10.1080/15298868.2012.681118}
\showDOI{\tempurl}


\bibitem[Potla and Potla(2025)]%
        {potla2025ai}
\bibfield{author}{\bibinfo{person}{Ravi~Teja Potla} {and} \bibinfo{person}{Ravi~Teja Potla}.} \bibinfo{year}{2025}\natexlab{}.
\newblock \showarticletitle{Ai-powered threat detection in online communities: A multi-modal deep learning approach}.
\newblock \bibinfo{journal}{\emph{Journal of Computer and Communications}} \bibinfo{volume}{13}, \bibinfo{number}{2} (\bibinfo{year}{2025}), \bibinfo{pages}{155--171}.
\newblock


\bibitem[Quayson(2007)]%
        {quayson2007aesthetic}
\bibfield{author}{\bibinfo{person}{Ato Quayson}.} \bibinfo{year}{2007}\natexlab{}.
\newblock \bibinfo{booktitle}{\emph{Aesthetic nervousness: Disability and the crisis of representation}}.
\newblock \bibinfo{publisher}{Columbia University Press}.
\newblock


\bibitem[{Rec Room}(nd)]%
        {RecRoom_CodeOfConduct}
\bibfield{author}{\bibinfo{person}{{Rec Room}}.} \bibinfo{year}{n.d.}\natexlab{}.
\newblock \bibinfo{title}{Code of Conduct}.
\newblock \bibinfo{howpublished}{\url{https://recroom.com/code-of-conduct}}.
\newblock
\newblock
\shownote{Accessed September 11, 2025}.


\bibitem[Sabri et~al\mbox{.}(2023)]%
        {sabri2023challenges}
\bibfield{author}{\bibinfo{person}{Nazanin Sabri}, \bibinfo{person}{Bella Chen}, \bibinfo{person}{Annabelle Teoh}, \bibinfo{person}{Steven~P Dow}, \bibinfo{person}{Kristen Vaccaro}, {and} \bibinfo{person}{Mai Elsherief}.} \bibinfo{year}{2023}\natexlab{}.
\newblock \showarticletitle{Challenges of moderating social virtual reality}. In \bibinfo{booktitle}{\emph{Proceedings of the 2023 CHI Conference on Human Factors in Computing Systems}}. \bibinfo{pages}{1--20}.
\newblock


\bibitem[Sanders(2006)]%
        {sanders2006overprotection}
\bibfield{author}{\bibinfo{person}{Katrina~Y Sanders}.} \bibinfo{year}{2006}\natexlab{}.
\newblock \showarticletitle{Overprotection and lowered expectations of persons with disabilities: The unforeseen consequences}.
\newblock \bibinfo{journal}{\emph{Work}} \bibinfo{volume}{27}, \bibinfo{number}{2} (\bibinfo{year}{2006}), \bibinfo{pages}{181--188}.
\newblock


\bibitem[Sannon et~al\mbox{.}(2023)]%
        {sannon2023disability}
\bibfield{author}{\bibinfo{person}{Shruti Sannon}, \bibinfo{person}{Jordyn Young}, \bibinfo{person}{Erica Shusas}, {and} \bibinfo{person}{Andrea Forte}.} \bibinfo{year}{2023}\natexlab{}.
\newblock \showarticletitle{Disability Activism on Social Media: Sociotechnical Challenges in the Pursuit of Visibility}. In \bibinfo{booktitle}{\emph{Proceedings of the 2023 CHI Conference on Human Factors in Computing Systems}} \emph{(\bibinfo{series}{CHI '23})}. \bibinfo{publisher}{Association for Computing Machinery}, \bibinfo{address}{New York, NY, USA}, \bibinfo{numpages}{15}~pages.
\newblock
\urldef\tempurl%
\url{https://doi.org/10.1145/3544548.3581333}
\showDOI{\tempurl}


\bibitem[Schafer et~al\mbox{.}(2023)]%
        {schafer2023participatory}
\bibfield{author}{\bibinfo{person}{Joseph~S Schafer}, \bibinfo{person}{Kate Starbird}, {and} \bibinfo{person}{Daniela~K Rosner}.} \bibinfo{year}{2023}\natexlab{}.
\newblock \showarticletitle{Participatory Design and Power in Misinformation, Disinformation, and Online Hate Research}. In \bibinfo{booktitle}{\emph{Proceedings of the 2023 ACM Designing Interactive Systems Conference}}. \bibinfo{pages}{1724--1739}.
\newblock


\bibitem[Schulenberg et~al\mbox{.}(2023a)]%
        {creepy_Schulenberg_2023}
\bibfield{author}{\bibinfo{person}{Kelsea Schulenberg}, \bibinfo{person}{Guo Freeman}, \bibinfo{person}{Lingyuan Li}, {and} \bibinfo{person}{Catherine Barwulor}.} \bibinfo{year}{2023}\natexlab{a}.
\newblock \showarticletitle{"Creepy Towards My Avatar Body, Creepy Towards My Body": How Women Experience and Manage Harassment Risks in Social Virtual Reality}.
\newblock \bibinfo{journal}{\emph{Proc. ACM Hum.-Comput. Interact.}} \bibinfo{volume}{7}, \bibinfo{number}{CSCW2}, Article \bibinfo{articleno}{236} (\bibinfo{date}{Oct.} \bibinfo{year}{2023}), \bibinfo{numpages}{29}~pages.
\newblock
\urldef\tempurl%
\url{https://doi.org/10.1145/3610027}
\showDOI{\tempurl}


\bibitem[Schulenberg et~al\mbox{.}(2023b)]%
        {Schulenberg_AImod_2023}
\bibfield{author}{\bibinfo{person}{Kelsea Schulenberg}, \bibinfo{person}{Lingyuan Li}, \bibinfo{person}{Guo Freeman}, \bibinfo{person}{Samaneh Zamanifard}, {and} \bibinfo{person}{Nathan~J. McNeese}.} \bibinfo{year}{2023}\natexlab{b}.
\newblock \showarticletitle{Towards Leveraging AI-based Moderation to Address Emergent Harassment in Social Virtual Reality}. In \bibinfo{booktitle}{\emph{Proceedings of the 2023 CHI Conference on Human Factors in Computing Systems}} (Hamburg, Germany) \emph{(\bibinfo{series}{CHI '23})}. \bibinfo{publisher}{Association for Computing Machinery}, \bibinfo{address}{New York, NY, USA}, Article \bibinfo{articleno}{514}, \bibinfo{numpages}{17}~pages.
\newblock
\showISBNx{9781450394215}
\urldef\tempurl%
\url{https://doi.org/10.1145/3544548.3581090}
\showDOI{\tempurl}


\bibitem[Schulenberg et~al\mbox{.}(2023c)]%
        {schulenberg_2023_birdcage}
\bibfield{author}{\bibinfo{person}{Kelsea Schulenberg}, \bibinfo{person}{Lingyuan Li}, \bibinfo{person}{Caitlin Lancaster}, \bibinfo{person}{Douglas Zytko}, {and} \bibinfo{person}{Guo Freeman}.} \bibinfo{year}{2023}\natexlab{c}.
\newblock \showarticletitle{" We Don't Want a Bird Cage, We Want Guardrails": Understanding \& Designing for Preventing Interpersonal Harm in Social VR through the Lens of Consent}.
\newblock \bibinfo{journal}{\emph{Proceedings of the ACM on Human-Computer Interaction}} \bibinfo{volume}{7}, \bibinfo{number}{CSCW2} (\bibinfo{year}{2023}), \bibinfo{pages}{1--30}.
\newblock


\bibitem[Shakespeare(1999)]%
        {shakespeare1999joking}
\bibfield{author}{\bibinfo{person}{Tom Shakespeare}.} \bibinfo{year}{1999}\natexlab{}.
\newblock \showarticletitle{Joking a part}.
\newblock \bibinfo{journal}{\emph{Body \& Society}} \bibinfo{volume}{5}, \bibinfo{number}{4} (\bibinfo{year}{1999}), \bibinfo{pages}{47--52}.
\newblock


\bibitem[Sharma(2022)]%
        {sharma_2022_personalboundary}
\bibfield{author}{\bibinfo{person}{Vivek Sharma}.} \bibinfo{year}{2022}\natexlab{}.
\newblock \showarticletitle{Introducing a Personal Boundary for {Horizon} Worlds and Venues}.
\newblock \bibinfo{journal}{\emph{Meta Newsroom}} (\bibinfo{date}{4 February} \bibinfo{year}{2022}).
\newblock
\urldef\tempurl%
\url{https://about.fb.com/news/2022/02/personal-boundary-horizon/}
\showURL{%
\tempurl}


\bibitem[Stetten et~al\mbox{.}(2019)]%
        {stetten2019SupportGroups}
\bibfield{author}{\bibinfo{person}{Nichole~E. Stetten}, \bibinfo{person}{Kelsea LeBeau}, \bibinfo{person}{Maria~A. Aguirre}, \bibinfo{person}{Alexis~B. Vogt}, \bibinfo{person}{Jazmine~R. Quintana}, \bibinfo{person}{Alexis~R. Jennings}, {and} \bibinfo{person}{Mark Hart}.} \bibinfo{year}{2019}\natexlab{}.
\newblock \showarticletitle{Analyzing the Communication Interchange of Individuals With Disabilities Utilizing Facebook, Discussion Forums, and Chat Rooms: Qualitative Content Analysis of Online Disabilities Support Groups}.
\newblock \bibinfo{journal}{\emph{JMIR Rehabilitation and Assistive Technologies}} \bibinfo{volume}{6}, \bibinfo{number}{2} (\bibinfo{year}{2019}), \bibinfo{pages}{e12667}.
\newblock
\urldef\tempurl%
\url{https://doi.org/10.2196/12667}
\showDOI{\tempurl}


\bibitem[Terry et~al\mbox{.}(2017)]%
        {terry2017thematic}
\bibfield{author}{\bibinfo{person}{Gareth Terry}, \bibinfo{person}{Nikki Hayfield}, \bibinfo{person}{Victoria Clarke}, \bibinfo{person}{Virginia Braun}, {et~al\mbox{.}}} \bibinfo{year}{2017}\natexlab{}.
\newblock \showarticletitle{Thematic analysis}.
\newblock \bibinfo{journal}{\emph{The SAGE handbook of qualitative research in psychology}} \bibinfo{volume}{2}, \bibinfo{number}{17-37} (\bibinfo{year}{2017}), \bibinfo{pages}{25}.
\newblock


\bibitem[Tough et~al\mbox{.}(2017)]%
        {tough2017social}
\bibfield{author}{\bibinfo{person}{Hannah Tough}, \bibinfo{person}{Johannes Siegrist}, {and} \bibinfo{person}{Christine Fekete}.} \bibinfo{year}{2017}\natexlab{}.
\newblock \showarticletitle{Social relationships, mental health and wellbeing in physical disability: a systematic review}.
\newblock \bibinfo{journal}{\emph{BMC public health}} \bibinfo{volume}{17}, \bibinfo{number}{1} (\bibinfo{year}{2017}), \bibinfo{pages}{414}.
\newblock


\bibitem[{VRChat Inc.}(2024)]%
        {VRChat_Community_Guidelines_2024}
\bibfield{author}{\bibinfo{person}{{VRChat Inc.}}} \bibinfo{year}{2024}\natexlab{}.
\newblock \bibinfo{title}{Community Guidelines}.
\newblock \bibinfo{howpublished}{\url{https://hello.vrchat.com/community-guidelines}}.
\newblock
\newblock
\shownote{Last updated August 12, 2024. Accessed September 11, 2025}.


\bibitem[{VRChat Inc.}(2025)]%
        {vrchat_safety_trust_2025}
\bibfield{author}{\bibinfo{person}{{VRChat Inc.}}} \bibinfo{year}{2025}\natexlab{}.
\newblock \bibinfo{title}{VRChat Safety and Trust System}.
\newblock \bibinfo{howpublished}{\url{https://docs.vrchat.com/docs/vrchat-safety-and-trust-system}}.
\newblock
\newblock
\shownote{Accessed: 2025-09-12}.


\bibitem[Wang et~al\mbox{.}(2015)]%
        {wang2015independent}
\bibfield{author}{\bibinfo{person}{Katie Wang}, \bibinfo{person}{Arielle Silverman}, \bibinfo{person}{Jason~D Gwinn}, {and} \bibinfo{person}{John~F Dovidio}.} \bibinfo{year}{2015}\natexlab{}.
\newblock \showarticletitle{Independent or ungrateful? Consequences of confronting patronizing help for people with disabilities}.
\newblock \bibinfo{journal}{\emph{Group Processes \& Intergroup Relations}} \bibinfo{volume}{18}, \bibinfo{number}{4} (\bibinfo{year}{2015}), \bibinfo{pages}{489--503}.
\newblock


\bibitem[Wang et~al\mbox{.}(2024)]%
        {wang2024hardenvr}
\bibfield{author}{\bibinfo{person}{Na Wang}, \bibinfo{person}{Jin Zhou}, \bibinfo{person}{Jie Li}, \bibinfo{person}{Bo Han}, \bibinfo{person}{Fei Li}, {and} \bibinfo{person}{Songqing Chen}.} \bibinfo{year}{2024}\natexlab{}.
\newblock \showarticletitle{HardenVR: Harassment Detection in Social Virtual Reality}. In \bibinfo{booktitle}{\emph{Proceedings of the 2024 IEEE Conference on Virtual Reality and 3D User Interfaces (IEEE VR)}}. \bibinfo{publisher}{IEEE}.
\newblock
\urldef\tempurl%
\url{https://doi.org/10.1109/VR58804.2024.00033}
\showDOI{\tempurl}


\bibitem[Weerasinghe et~al\mbox{.}(2025)]%
        {weerasinghe2025beyond}
\bibfield{author}{\bibinfo{person}{Maheshya Weerasinghe}, \bibinfo{person}{Shaun Macdonald}, \bibinfo{person}{Cristina Fiani}, \bibinfo{person}{Joseph O'Hagan}, \bibinfo{person}{Mathieu Chollet}, \bibinfo{person}{Mark McGill}, {and} \bibinfo{person}{Mohamed Khamis}.} \bibinfo{year}{2025}\natexlab{}.
\newblock \showarticletitle{Beyond mute and block: adoption and effectiveness of safety tools in social VR, from ubiquitous harassment to social sculpting}.
\newblock \bibinfo{journal}{\emph{IEEE Transactions on Visualization and Computer Graphics}} (\bibinfo{year}{2025}).
\newblock


\bibitem[Williamson et~al\mbox{.}(2021)]%
        {williamson2021proxemics}
\bibfield{author}{\bibinfo{person}{Julie Williamson}, \bibinfo{person}{Jie Li}, \bibinfo{person}{Vinoba Vinayagamoorthy}, \bibinfo{person}{David~A Shamma}, {and} \bibinfo{person}{Pablo Cesar}.} \bibinfo{year}{2021}\natexlab{}.
\newblock \showarticletitle{Proxemics and social interactions in an instrumented virtual reality workshop}. In \bibinfo{booktitle}{\emph{Proceedings of the 2021 CHI conference on human factors in computing systems}}. \bibinfo{pages}{1--13}.
\newblock


\bibitem[Wobbrock et~al\mbox{.}(2011)]%
        {wobbrock2011aligned}
\bibfield{author}{\bibinfo{person}{Jacob~O Wobbrock}, \bibinfo{person}{Leah Findlater}, \bibinfo{person}{Darren Gergle}, {and} \bibinfo{person}{James~J Higgins}.} \bibinfo{year}{2011}\natexlab{}.
\newblock \showarticletitle{The aligned rank transform for nonparametric factorial analyses using only anova procedures}. In \bibinfo{booktitle}{\emph{Proceedings of the SIGCHI conference on human factors in computing systems}}. \bibinfo{pages}{143--146}.
\newblock


\bibitem[Xu et~al\mbox{.}(2024)]%
        {xu2024safe}
\bibfield{author}{\bibinfo{person}{Yiwen Xu}, \bibinfo{person}{Qinyang Hou}, \bibinfo{person}{Hongyu Wan}, {and} \bibinfo{person}{Mirjana Prpa}.} \bibinfo{year}{2024}\natexlab{}.
\newblock \showarticletitle{Safe guard: an llm-agent for real-time voice-based hate speech detection in social virtual reality}.
\newblock \bibinfo{journal}{\emph{arXiv preprint arXiv:2409.15623}} (\bibinfo{year}{2024}).
\newblock


\bibitem[Young(2014)]%
        {young2014m}
\bibfield{author}{\bibinfo{person}{Stella Young}.} \bibinfo{year}{2014}\natexlab{}.
\newblock \showarticletitle{I’m not your inspiration, thank you very much}.
\newblock \bibinfo{journal}{\emph{TED: Ideas Worth Spreading. https://www. ted. com/talks/stella\_young\_i\_m\_not\_your\_inspi ration\_thank\_you\_very\_much}} (\bibinfo{year}{2014}).
\newblock


\bibitem[Zamanifard and Freeman(2019)]%
        {zamanifard2019togetherness}
\bibfield{author}{\bibinfo{person}{Samaneh Zamanifard} {and} \bibinfo{person}{Guo Freeman}.} \bibinfo{year}{2019}\natexlab{}.
\newblock \showarticletitle{" The togetherness that we crave" experiencing social VR in long distance relationships}. In \bibinfo{booktitle}{\emph{Companion Publication of the 2019 Conference on Computer Supported Cooperative Work and Social Computing}}. \bibinfo{pages}{438--442}.
\newblock


\bibitem[Zhang et~al\mbox{.}(2022)]%
        {zhang_2022}
\bibfield{author}{\bibinfo{person}{Kexin Zhang}, \bibinfo{person}{Elmira Deldari}, \bibinfo{person}{Zhicong Lu}, \bibinfo{person}{Yaxing Yao}, {and} \bibinfo{person}{Yuhang Zhao}.} \bibinfo{year}{2022}\natexlab{}.
\newblock \showarticletitle{“It’s Just Part of Me:” Understanding Avatar Diversity and Self-presentation of People with Disabilities in Social Virtual Reality}. In \bibinfo{booktitle}{\emph{Proceedings of the 24th International ACM SIGACCESS Conference on Computers and Accessibility}} (Athens, Greece) \emph{(\bibinfo{series}{ASSETS '22})}. \bibinfo{publisher}{Association for Computing Machinery}, \bibinfo{address}{New York, NY, USA}, Article \bibinfo{articleno}{4}, \bibinfo{numpages}{16}~pages.
\newblock
\showISBNx{9781450392587}
\urldef\tempurl%
\url{https://doi.org/10.1145/3517428.3544829}
\showDOI{\tempurl}


\bibitem[Zhang et~al\mbox{.}(2023)]%
        {zhang_2023_diary}
\bibfield{author}{\bibinfo{person}{Kexin Zhang}, \bibinfo{person}{Elmira Deldari}, \bibinfo{person}{Yaxing Yao}, {and} \bibinfo{person}{Yuhang Zhao}.} \bibinfo{year}{2023}\natexlab{}.
\newblock \showarticletitle{A Diary Study in Social Virtual Reality: Impact of Avatars with Disability Signifiers on the Social Experiences of People with Disabilities}. In \bibinfo{booktitle}{\emph{Proceedings of the 25th International ACM SIGACCESS Conference on Computers and Accessibility}} (New York, NY, USA) \emph{(\bibinfo{series}{ASSETS '23})}. \bibinfo{publisher}{Association for Computing Machinery}, \bibinfo{address}{New York, NY, USA}, Article \bibinfo{articleno}{40}, \bibinfo{numpages}{17}~pages.
\newblock
\showISBNx{9798400702204}
\urldef\tempurl%
\url{https://doi.org/10.1145/3597638.3608388}
\showDOI{\tempurl}


\bibitem[Zhang et~al\mbox{.}(2025)]%
        {zhang_2025_guidelines}
\bibfield{author}{\bibinfo{person}{Kexin Zhang}, \bibinfo{person}{Jr. Spencer, Edward Glenn~Scott}, \bibinfo{person}{Abijith Manikandan}, \bibinfo{person}{Andric Li}, \bibinfo{person}{Ang Li}, \bibinfo{person}{Yaxing Yao}, {and} \bibinfo{person}{Yuhang Zhao}.} \bibinfo{year}{2025}\natexlab{}.
\newblock \showarticletitle{Inclusive Avatar Guidelines for People with Disabilities: Supporting Disability Representation in Social Virtual Reality}. In \bibinfo{booktitle}{\emph{Proceedings of the 2025 CHI Conference on Human Factors in Computing Systems}} \emph{(\bibinfo{series}{CHI '25})}. \bibinfo{publisher}{Association for Computing Machinery}, \bibinfo{address}{New York, NY, USA}, Article \bibinfo{articleno}{560}, \bibinfo{numpages}{26}~pages.
\newblock
\showISBNx{9798400713941}
\urldef\tempurl%
\url{https://doi.org/10.1145/3706598.3714230}
\showDOI{\tempurl}


\bibitem[Zheng et~al\mbox{.}(2023)]%
        {zheng2023understanding}
\bibfield{author}{\bibinfo{person}{Qingxiao Zheng}, \bibinfo{person}{Shengyang Xu}, \bibinfo{person}{Lingqing Wang}, \bibinfo{person}{Yiliu Tang}, \bibinfo{person}{Rohan~C Salvi}, \bibinfo{person}{Guo Freeman}, {and} \bibinfo{person}{Yun Huang}.} \bibinfo{year}{2023}\natexlab{}.
\newblock \showarticletitle{Understanding safety risks and safety design in social VR environments}.
\newblock \bibinfo{journal}{\emph{Proceedings of the ACM on Human-Computer Interaction}} \bibinfo{volume}{7}, \bibinfo{number}{CSCW1} (\bibinfo{year}{2023}), \bibinfo{pages}{1--37}.
\newblock


\end{thebibliography}
